\documentclass[%
 reprint,
 amsmath,amssymb,
 aps,
 prd,
floatfix,
showkeys
]{revtex4-1}

\usepackage[USenglish]{babel}
\usepackage{slashed}
\usepackage{graphicx}
\usepackage{dcolumn}
\usepackage{bm}
\usepackage{amssymb}
\usepackage{amsmath}
\usepackage{graphicx}
\usepackage{epsfig}
\usepackage{rotating}

\usepackage[utf8]{inputenc} 
\usepackage{amsmath} 
\usepackage{array}
\usepackage{float}
\usepackage{color}
\usepackage[usenames,dvipsnames]{xcolor}
\usepackage{epstopdf}
\usepackage{natbib}
\usepackage{graphicx}
\usepackage{lineno}
\usepackage{slashed}
\usepackage{color}
\usepackage{hyperref}
\usepackage{float}
\usepackage{mathtools,slashed}
\usepackage{multirow}

\newcommand{\be}{\begin{equation}}
\newcommand{\ee}{\end{equation}}
\newcommand{\bea}{\begin{eqnarray}}
\newcommand{\eea}{\end{eqnarray}}
\newcommand{\beas}{\begin{eqnarray*}}
\newcommand{\eeas}{\end{eqnarray*}}

\newcommand{\bse}{\begin{subequations}}
\newcommand{\ese}{\end{subequations}}

\begin{document}

\title{Solar irradiance statistical analysis in Mexico City from 2018 to 2021}
\author{C. H.~Zepeda~Fern\'andez$^{1,2}$}
\email{Corresponding author, email:hzepeda@fcfm.buap.mx}
\author{M. Ram\'irez-Cer\'on$^{3}$}
\author{X. Cancino-Rojas$^{2}$}
\author{E. Moreno~Barbosa$^{2}$}
\author{M. Arredondo-Vel\'azquez$^{2}$}
\author{J. M. Her\'andez-L\'opez$^2$}

\address{
$^1$Cátedra CONACyT, 03940, CdMx Mexico\\
$^2$Facultad de Ciencias F\'isico Matem\'aticas, Benem\'erita Universidad Aut\'onoma de Puebla, Av. San Claudio y 18 Sur, Ciudad Universitaria 72570, Puebla, Mexico\\
$^3$Universidad de las América, Puebla
}

%%%%%% Abstract %%%%%%
\begin{abstract}
Solar radiation is made up of three components of electromagnetic waves: infrared, visible and ultraviolet. The infrared component is the cause of thermal energy, the visible spectrum allows to see through the eyes and the ultraviolet component is the most energetic and damaging. Solar radiation has several benefits, such as helping to synthesize vitamin D in the skin, favors blood circulation, among others benefits for the human body. In the Earth, it is the main source of energy for agriculture, also used as an alternative source of energy to hydrocarbons, through solar cells. The solar irradiance represents the surface power density with units W/m$^2$ in SI. Too much exposure can cause damage and an increase in value over the time can be can be also damaging. In this work it was used an open data base provided by Secretar\'ia del Medio Ambiente, from which a statistical analysis was performed of the solar irradiance values measured at various meteorological stations in Mexico City and the so-called metropolitan area, from 2018 to 2021. This analysis was carried out per years, months and days. From the solar irradiance values distributions, it was obtained the averages, maximums and means were it was found there was no variation in the solar irradiance values over this period of years.

\end{abstract}

\maketitle

%%%%%% Main Text %%%%%%

\section{Introduction}\label{introduction}
The solar irradiance is defined by the power per unit area (W/m$^2$, SI), it is conformed by electromagnetic waves, which are: 
\begin{enumerate}
\item Infrared radiation (IR): It is the cause of the thermal energy, which is classified by
\begin{itemize}
    \item Infrared type A (IRA), in a range of 780-1400~nm.
    \item Infrared type B (IRB), in a range of 1400-3000~nm.
    \item Infrared type C (IRC), in a range of 1~mm to 3000~nm.
\end{itemize} 
\item Visible radiation (Vi): It is the part of the electromagnetic spectrum through the human eye sees. It is in a range of 400-780~nm.
\item Ultraviolet radiation (UV): It is the most energetic part of the radiation. Similar to IR, the UV is classified by:
    \begin{itemize}
   \item UV type A (UVA), in a range of 315-400~nm.
    \item UV type B (UVB), in a range of 280-315~nm.
    \item UV type C (UVC), in a range of 100-280~nm.
    \end{itemize}
\end{enumerate}
UVC rays are the most energetic, which makes them the most dangerous for humanity. However, they are absorbed in the ozone layer, while  UVB and UVA rays are not fully braked~\cite{WHO}. It is well known that  solar irradiance is one of the factors for life on the Earth, it has some benefits, such as on the skin, which helps make vitamin D. Also it also helps agriculture and it can be used as an alternative source of energy used in solar cells. However, too much exposure can cause skin damage~\cite{damage1,damage2,damage3,damage4,damage5}, due to  UV and IR components, which can causes breakdown of collagen fibers in the skin. However, recently studies have shown that the right combination of IR and IUV can improve skin disorders~\cite{skincare}. In the case of objects receiving radiation daily, they tend to become opaque and lose its properties. To prevent this issues, it is defined a dimensionless  value for the UV component, which provides a prevention of sun exposure, it is called the ultraviolet index (IUV)~\cite{Filetov2010TheUI, WHO}. It normally is represented in a range from 1 to 11+, where the 11+ means that the upper limit may be greater than eleven. The World Health Organization classifies the values 1 and 2 as low; 3, 4 and 5 as moderate; 6 and 7 as high; 8, 9 and 10 very high and 11 and more values as extreme~\cite{WHO}. For IUV$\leq2$ one can go freely in the outdoors. For IUV values from 3 to 7, it is recommended to use sunscreen, hat and long sleeve shirt, as well as, seek shade at midday. Finally, for the case when the IUV$\geq8$ it is recommended to avoid being outside at midday and it is essential to use sunscreen, hat and long sleeve shirt.\\
As it was mentioned, the ozone layer is a protection to keep out the UVC, over the years it has decreased due to pollution, aerosol, etc. To deplete the ozone layer, the Mont Real protocol was created~\cite{MontReal}. Several studies have been done where it has been shown that the UV index has not been change~\cite{UV1,UV2}. Predictions have been made by simulations in order to study the UV index and the effective dose for the production of vitamin D in the human skin~\cite{C5PP00093A}, where it had shown an improvement due to the recovery of ozone layer. These simulations are based on angle radiation  over the Earth~\cite{ozone}.\\
Another relevant characteristic of the Sun is its magnetic field, which change its direction, this is called Solar cycle. It is related to the sunspots, which decreases and increases around eleven years~\cite{sunspots}. The effects on Earth due to the Solar cycle are reflected on change climate, radio frequency communications, navigation, among others. Studies carried out have shown that the Solar cycle occurs between ten and eleven years~\cite{ZHU20233521, unknown, articleChe}
%, SUN2016121, SolarCyclePrediction1, SolarCyclePrediction2}. 
It has been noted that the maximum of the Solar cycle have occurred at 1993, 2003 and 2013, it is expected that the next maximum value of the cycle occurs in 2025~\cite{2025}. These measurements have been a challenge, because solar irradiation on Earth varies due to pollution, urban development and aerosols~\cite{popullation1,popullation2,popullation3,popullation4,popullation5}. This causes fluctuations in measured values.\\
The solar irradiance variation studies have been of great importance since their have effects on climate. However, several studies have shown that despite oscillatory behavior, the solar irradiance has not shown increase in values~\cite{NASA}. Although the temperature has increased the radiation has remained constant, according to measurements tacking from 1880 to 2020~\cite{NASA2}. One of the current work of interest with solar irradiance variation, is to find a model that describes it~\cite{model, model2}, having as main application irrigation problems~\cite{article_irradiance1} or construction of buildings with low energy consumption~\cite{article_irradiance2}, where also, it has been shown that there have been no changes over the years.\\
In Mexico, around 16,000 new cases of skin cancer have been diagnosed every year, it has been estimated that skin cancer doubles every ten years~\cite{secretariasalud, economista}. Therefore the number of cases of melanoma has increased~\cite{dermatologia}, this disease  being the most skin dangerous and lethal~\cite{gallegos}. It mainly affects farmers, fishermen, street vendors, etc. Which can also cause eye damage~\cite{mata}.\\

In this work, it was made a solar irradiance statistical analysis, where it was used an open data base provided by the \textit{Secretar\'ia del Medio Ambiente} (SEDEMA), this data was acquired by various weather stations in Mexico City and the so-called Metropolitan Area. The database information of SEDEMA is explained in Section~\ref{DataAdcquisition}. In Section~\ref{AnalysisResults} is shown the data analysis, in which the fluctuations in solar irradiance values distribution are shown in the period from 2018 to 2021. This analysis could be carried out for years (Subsection~\ref{Annual}), months (Subsection~\ref{Monthly}) and days (Subsection~\ref{Daily}). Finally, in Section~\ref{Discussions} the conclusions and discussions are shown.

\section{Open data base of SEDEMA}\label{DataAdcquisition}
\begin{figure}
    \centering
    \includegraphics[width=0.4\linewidth]{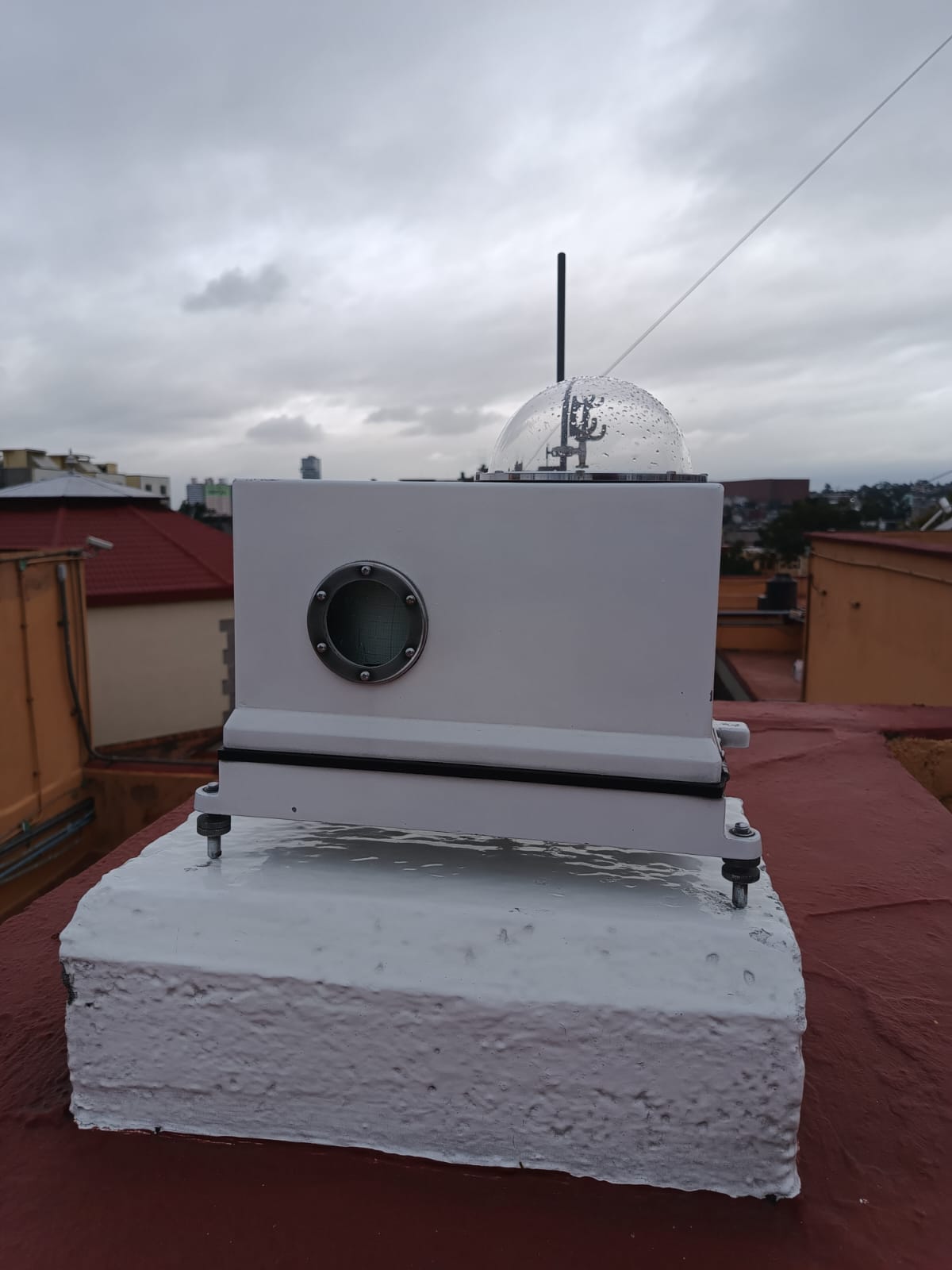}
        \includegraphics[width=0.4\linewidth]{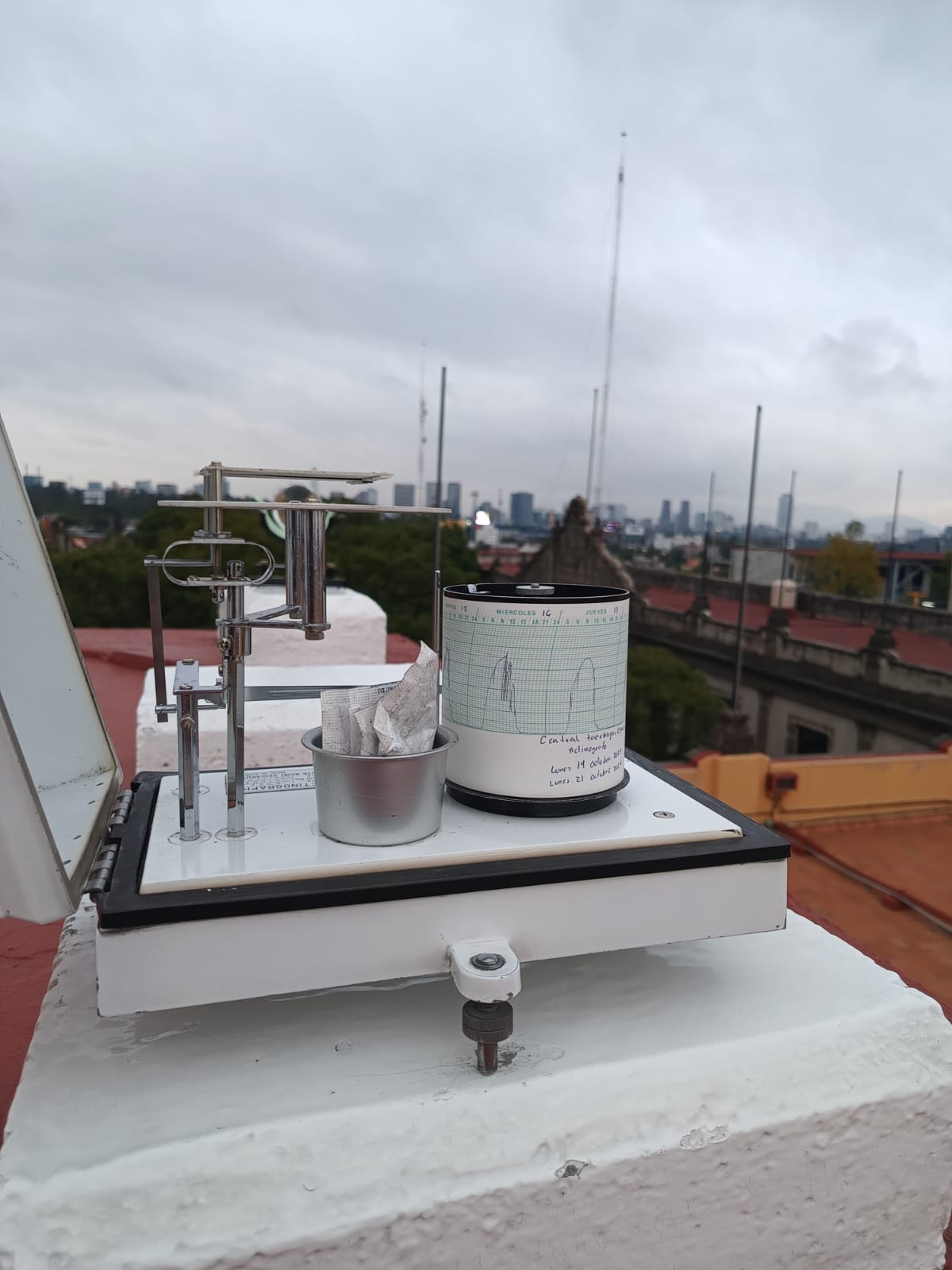}
    \caption{Actinograph used for solar irradiance measurement at the weather stations in CMDX.}
    \label{actinograph}
\end{figure}

The government of Mexico City (CDMX) has an open data base of solar irradiance values (SIV), which contains measurements during each hour of ultraviolet A radiation (UVA), from 2018 to 2021~\cite{base}. There are 44 monitoring stations, located in different locations of CDMX and the so-called Metropolitan area. The information provided is obtained from eleven stations:
\begin{itemize}
    \item Chalco (CHO)  
    \item Cuautitl\'an (CUT) 
    \item FES Actal\'an (FAC)
     \item Laboratorio de An\'alisis Ambiental (LAA)
    \item Merced (MER)
    \item Montencillo (MON)
    \item Milpa Alta (MPA)
    \item Pedregal (PED)
    \item San Agust\'in (SAG)
    \item Santa Fe (SFE)
    \item Tlalnepantla (TLA)
\end{itemize}
For each year,  data information are collected in a ``.xls" format and it is arranged in eight columns: Day, month, year, hour, name of the station, parameter name (which is UVA), SIV and the number 17, which represents the units (mW/cm$^2$), according to the manual~\cite{base}.\\
The data was taken using an actinograph elaborated by \textit{OTA KEIKE SEISAKUSHO}, which principal function is the solar radiation measurement, it is shown in Figure~\ref{actinograph}. %Finally, as a note, there is null data for certain hours during the days.

\section{Analysis and results}\label{AnalysisResults}
Approximately between eight in the evening and six in the morning (depending of the day of the year) there are zero values measured, this is because radiation does not reach to CDMX. However, outside this period, there are null values for several hours, which means no data was taken, for example, due to some technical problem. All SIV equal to zero were omitted for this analysis. As mentioned above,  measurements were taken over one-hour periods, therefore, it possible to make an analysis per years (Subsection~\ref{Annual}), months (Subsection~\ref{Monthly}) and days (Subsection~\ref{Daily}), as shown below.

\subsection{Annual solar radiation study}\label{Annual}
\begin{figure}
    \centering
    \includegraphics[width=0.6\linewidth]{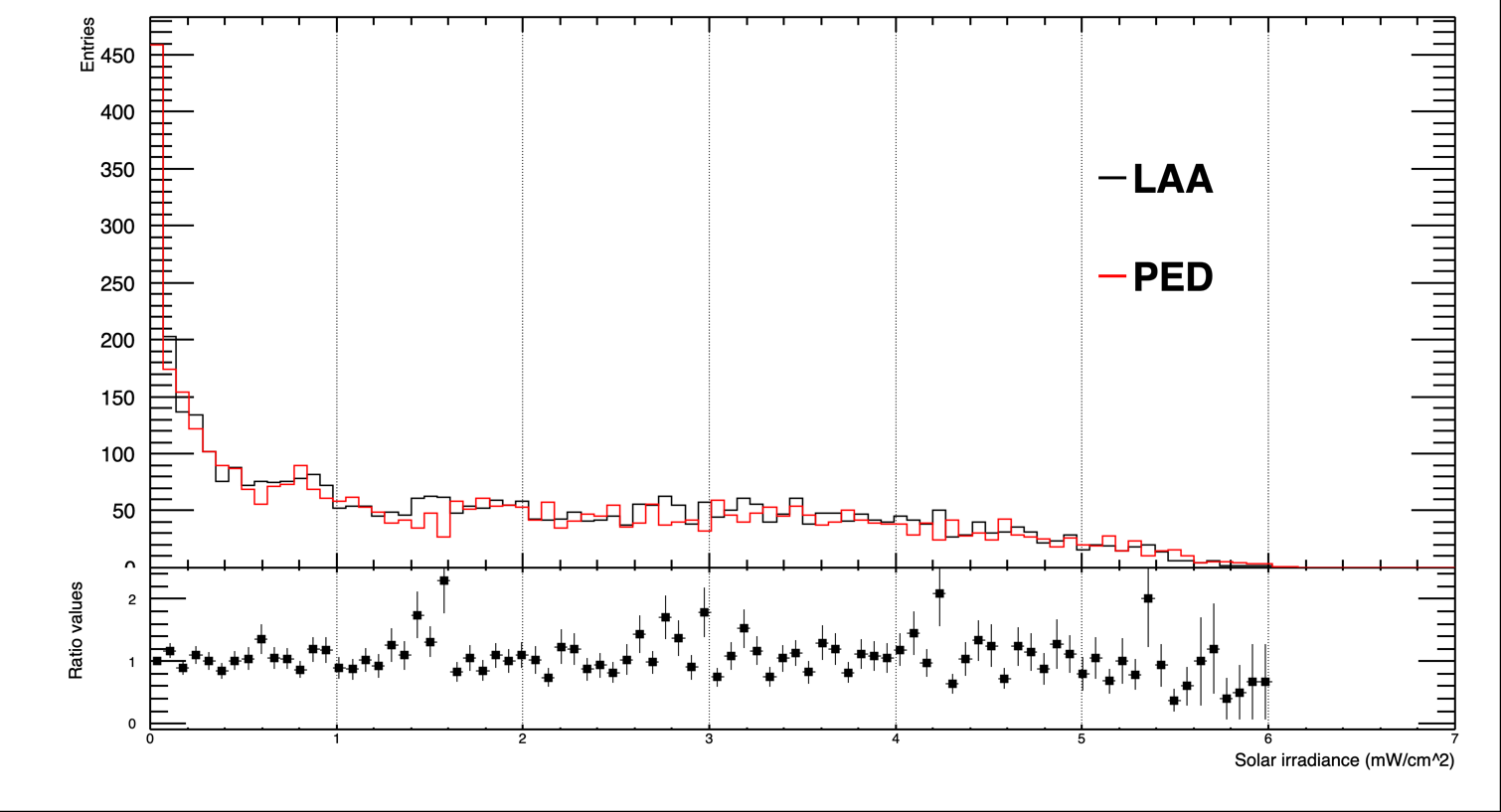}
       \includegraphics[width=0.6\linewidth]{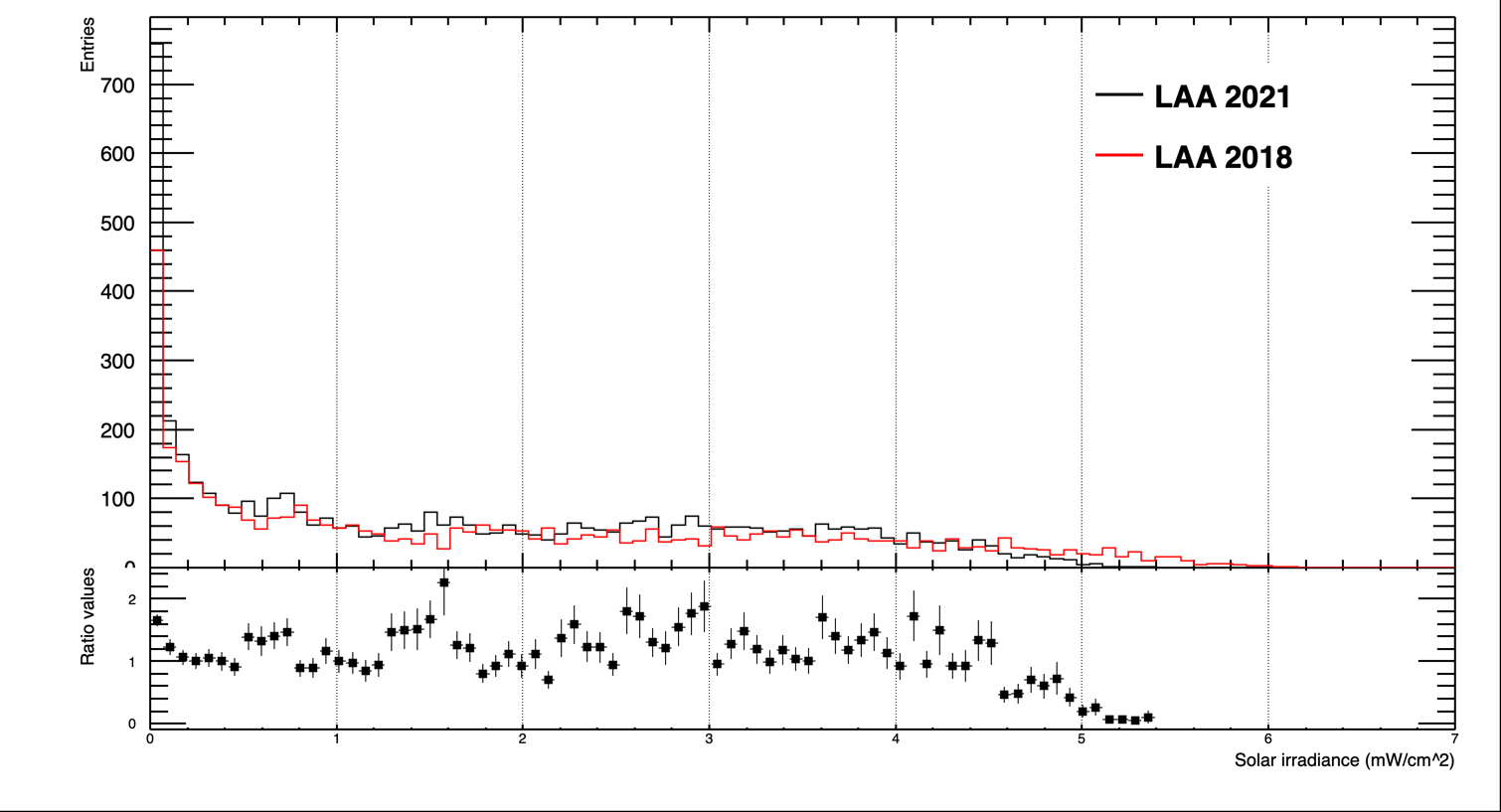}
    \caption{SIV distribution for  LAA (red) and PED (black) stations for 2018 (top) and LAA station for 2018 and 2021 (down).}
    \label{ratio_LAA_PED}
\end{figure}

As an example of the SIV annual distribution, in Figure~\ref{ratio_LAA_PED} it is shown for LAA and PED stations in 2018, where also it is shown the ratio between histograms (low panel). For these stations, it can be noted that the values of the ratio oscillates around one, both stations measured almost the same number of the events, being 4,218 events for PED and 4,498 events for LAA. Which means that the radiation measured annually is consistent at both stations. As it was mentioned above, there are null values, then a similar comparison shows a discrepancy between other stations and years, because the same events are not recorded. A similar analysis was performed focusing a station and comparing the SIV distribution between years, were similar results were obtained. As an example in Figure~\ref{ratio_LAA_PED} it is shown the SIV distribution for LAA station for 2018 and 2021.

\begin{figure}
    \centering
    \includegraphics[width=0.6\linewidth]{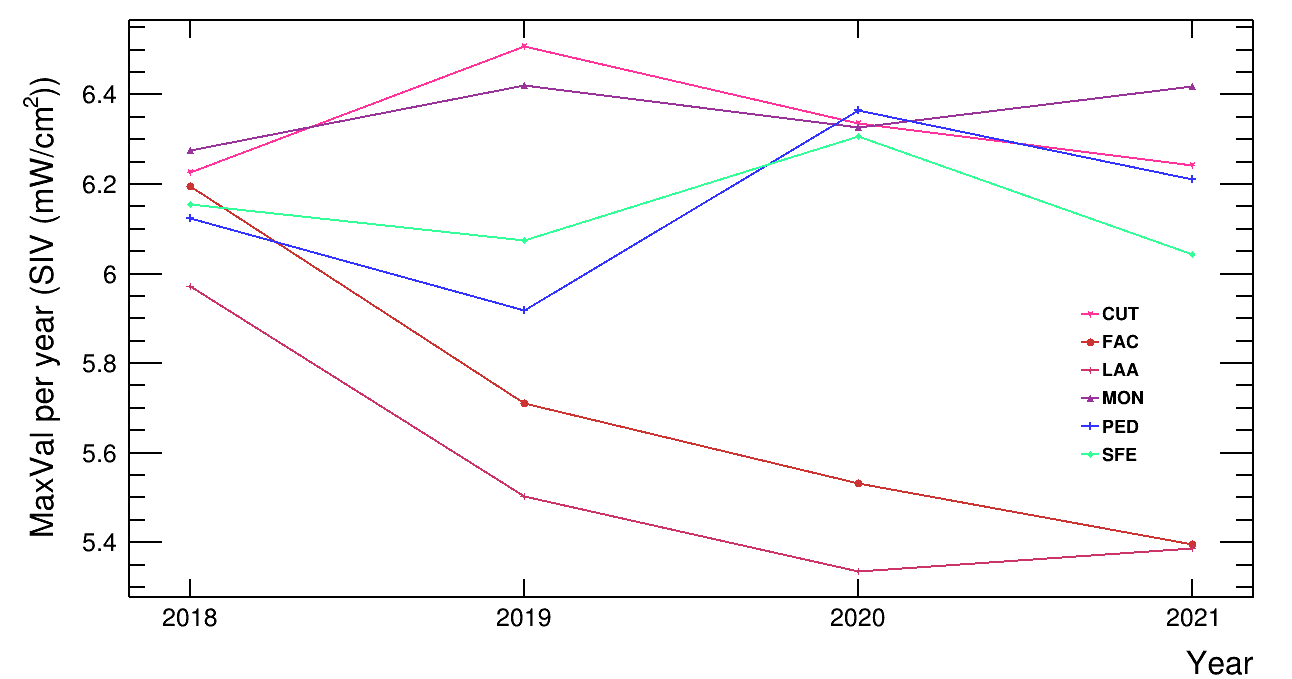}
    \includegraphics[width=0.6\linewidth]{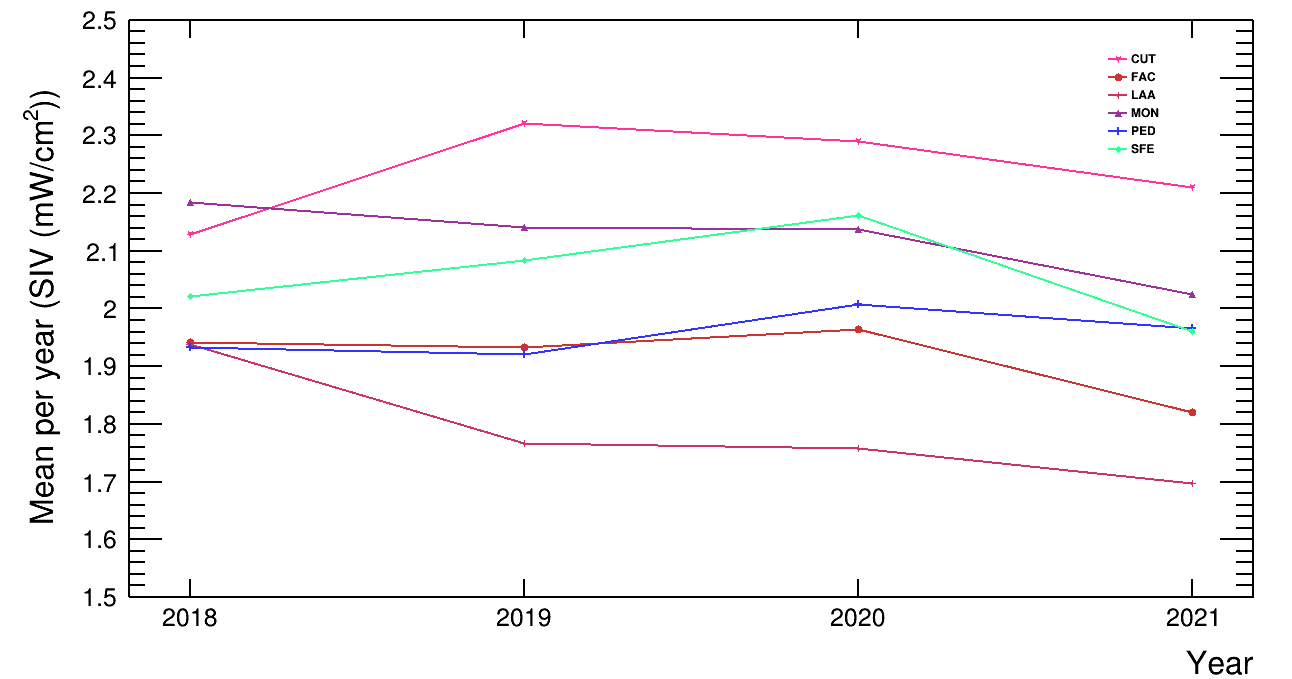}
    \caption{Maximum (top) and average (down) of SIV distribution for all stations over the years.}
    \label{annual_ave_max}
\end{figure}

\begin{figure}
    \centering
    \includegraphics[width=0.6\linewidth]{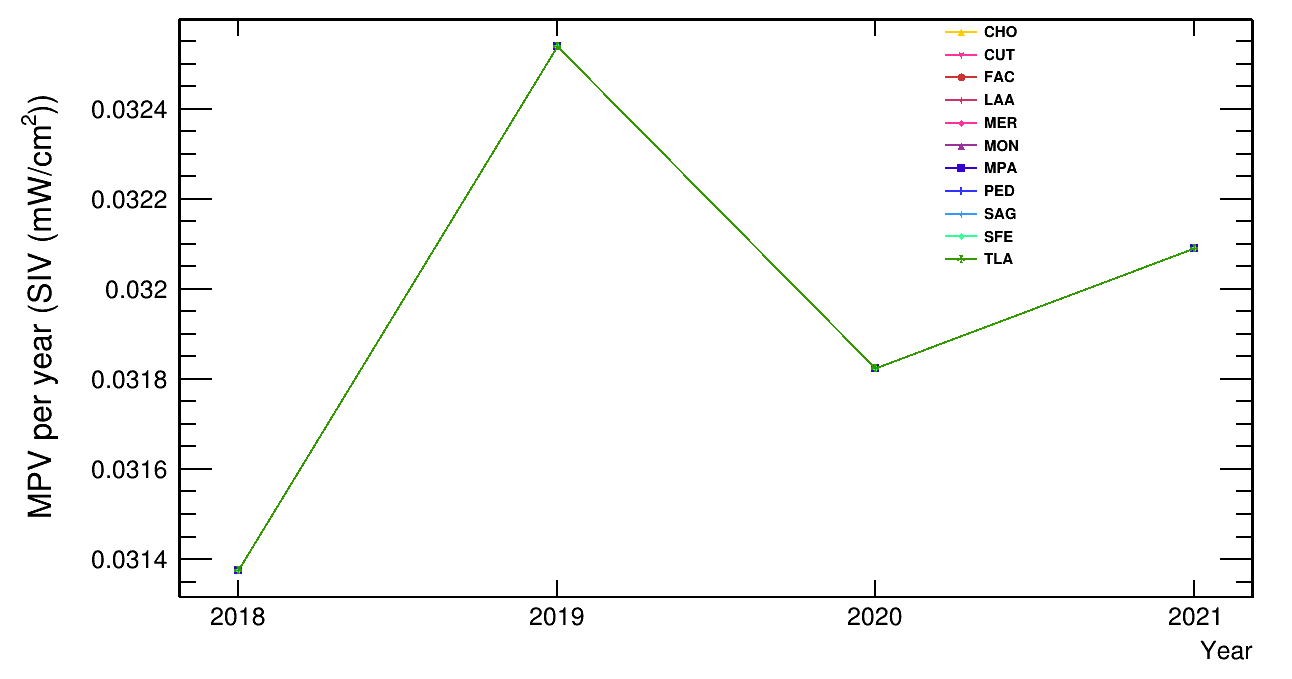}
    \caption{MPV ditribution for all stations and over the years, obtained from the SIV annual distribution. Graphs have the same MPV, which is why they are on top of each other.}
    \label{mpvvalueperyear}
\end{figure}

\begin{figure}
    \centering
    \includegraphics[width=0.6\linewidth]{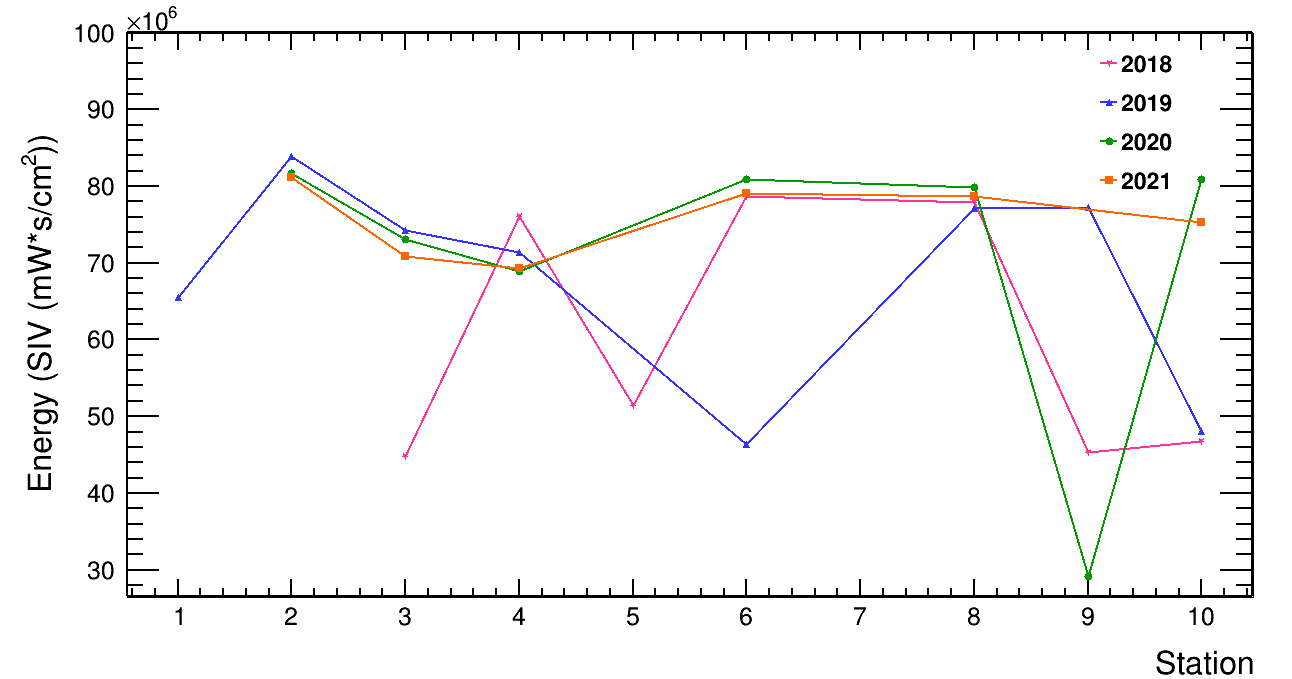}
    \caption{Total energy accumulated in LAA station, for each year.}
    \label{intyear}
\end{figure}

To conclude this annual analysis, it was obtained  the average and maximum of the SIV distributions, in Figure~\ref{annual_ave_max} it is shown the behavior of these values for six stations from 2018 to 2021. Only six stations are shown for a better appreciation. Slight fluctuations are observed, which are due to the null values.\\
The most probably value (MPV) also could be obtained from the SIV distribution. The behavior is shown in Figure~\ref{mpvvalueperyear} for all stations from 2018 to 2021, where the graphs are one on top of the other and having fluctuations of $10^{-3}~$mW/cm$^2$, therefore there has been no change on MPV. Finally, it was possible to compare the density energy (J/cm$^2$) by obtaining the integral of the SIV vs hour distribution for each station and year. The distribution is shown in Figure~\ref{intyear}, where the meteorological stations did not take data for several hours, which is why those drops in values.

\subsection{Monthly solar radiation study}\label{Monthly}
\begin{figure}
    \centering
    \includegraphics[width=0.6\linewidth]{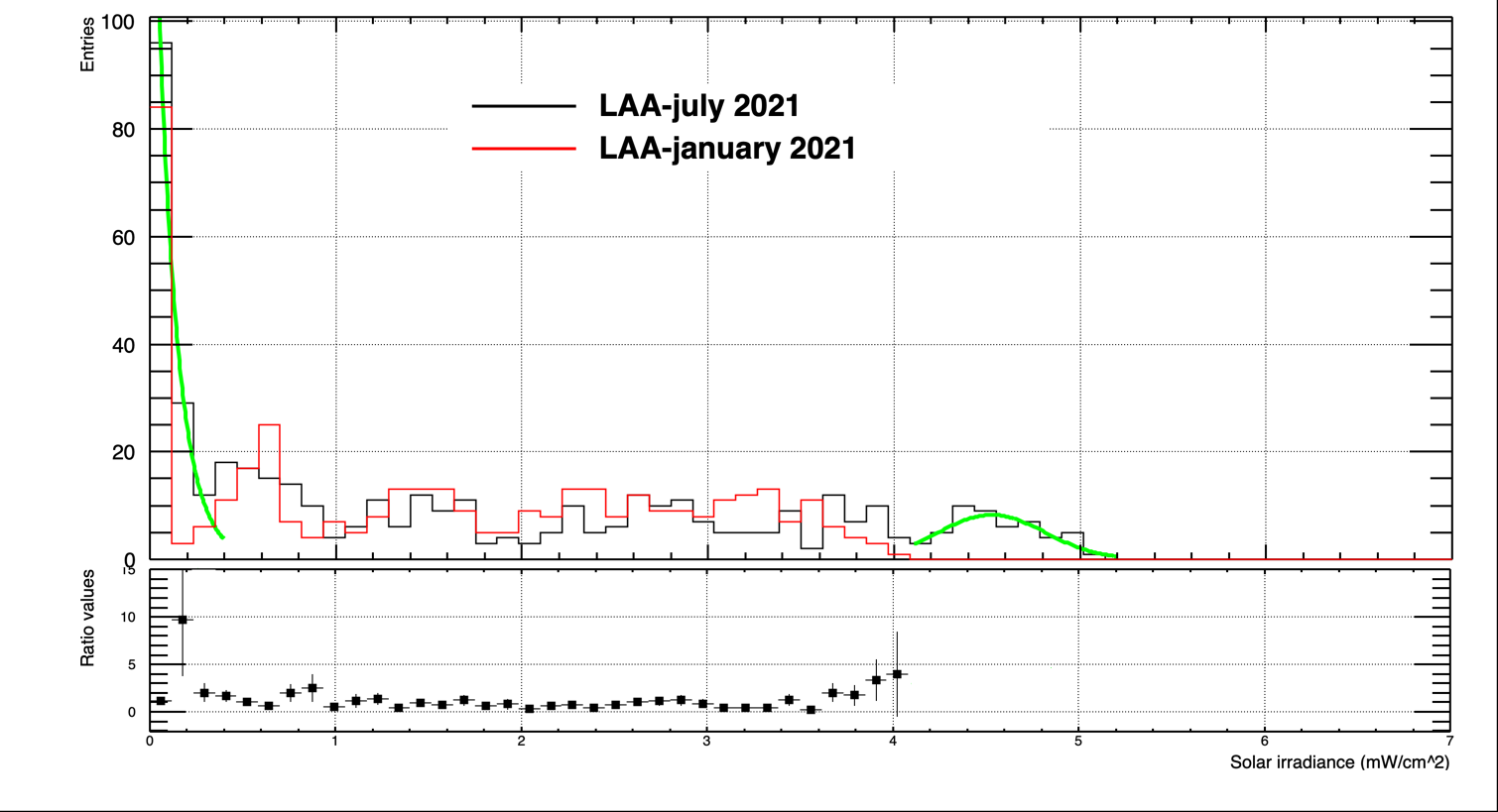}
    \caption{SIV distribution for the LAA station. It is shown the January (red) and July (black) distribution in 2021. The green lines shown an exponential fit for the fall of the first part  and the Gaussian fit for the last part of the distribution.}
    \label{His_jul_Jan}
\end{figure}

\begin{figure}
    \centering
    \includegraphics[width=0.6\linewidth]{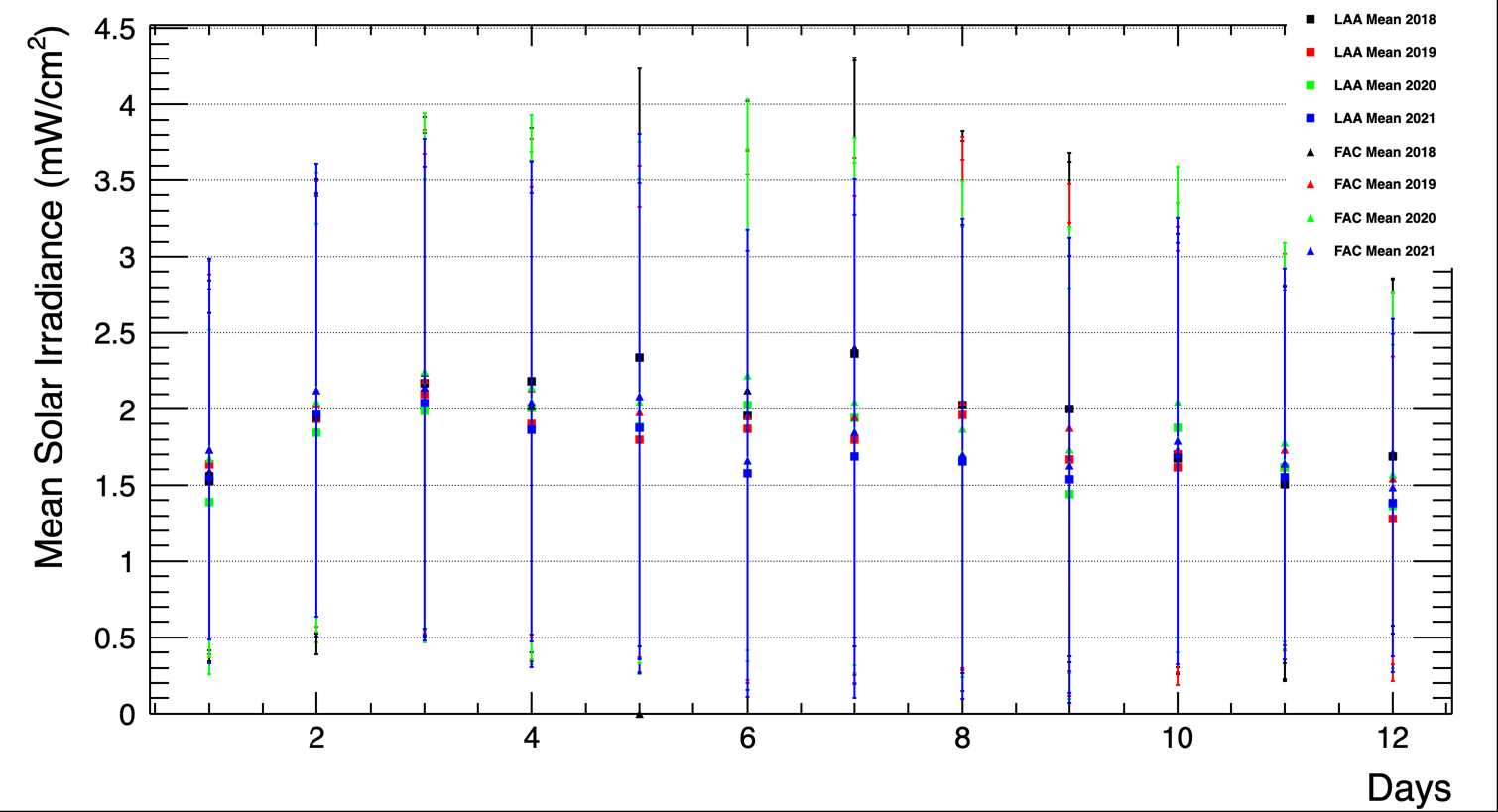}
    \includegraphics[width=0.6\linewidth]{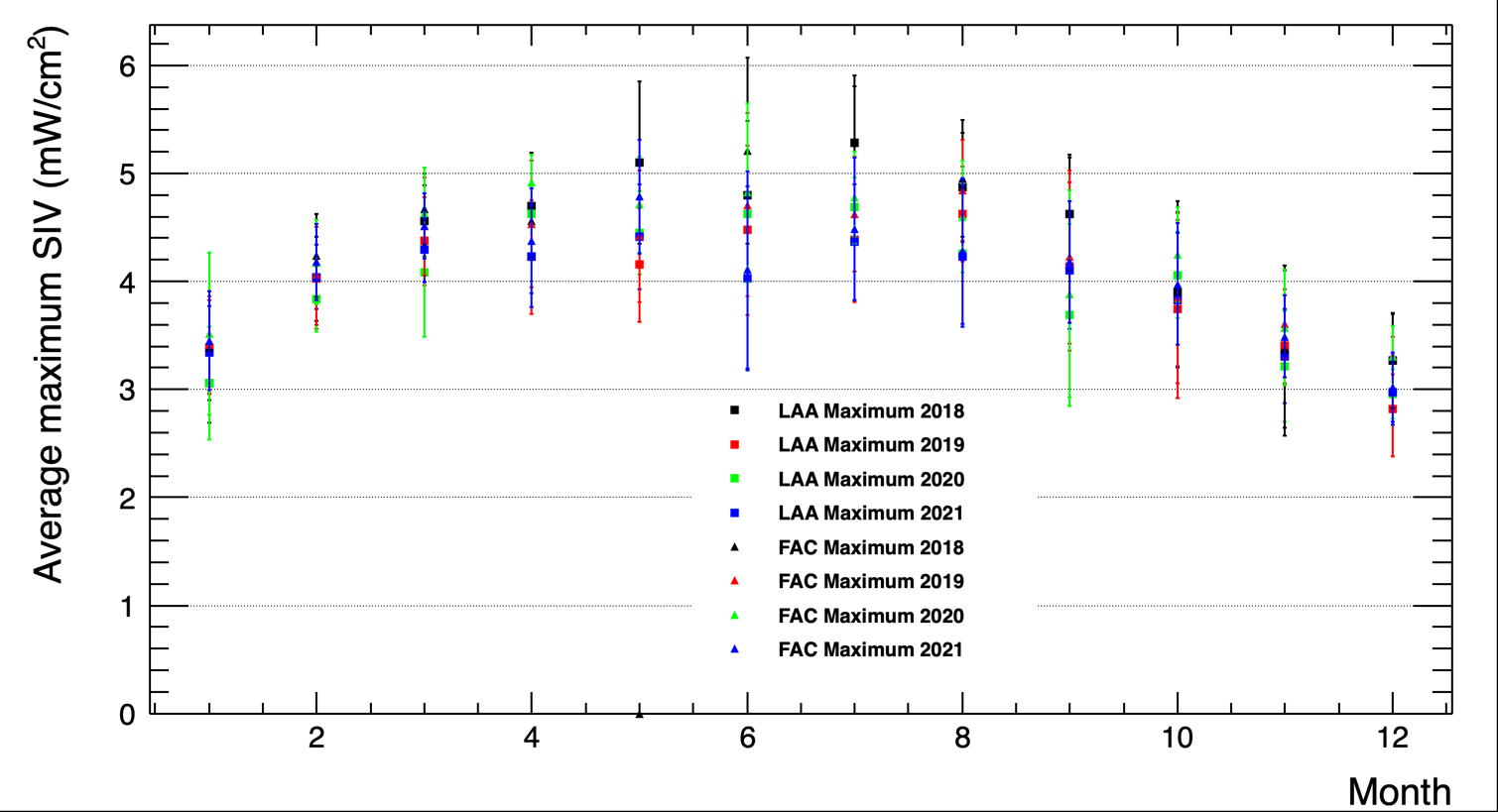}
    \caption{Mean (top) and maximum (down) average SIV distribution for the LAA and FAC stations. The distribution is plotted for each month and year.}
    \label{mean_LAA_FAC}
\end{figure}

\begin{figure}
    \centering
    \includegraphics[width=0.6\linewidth]{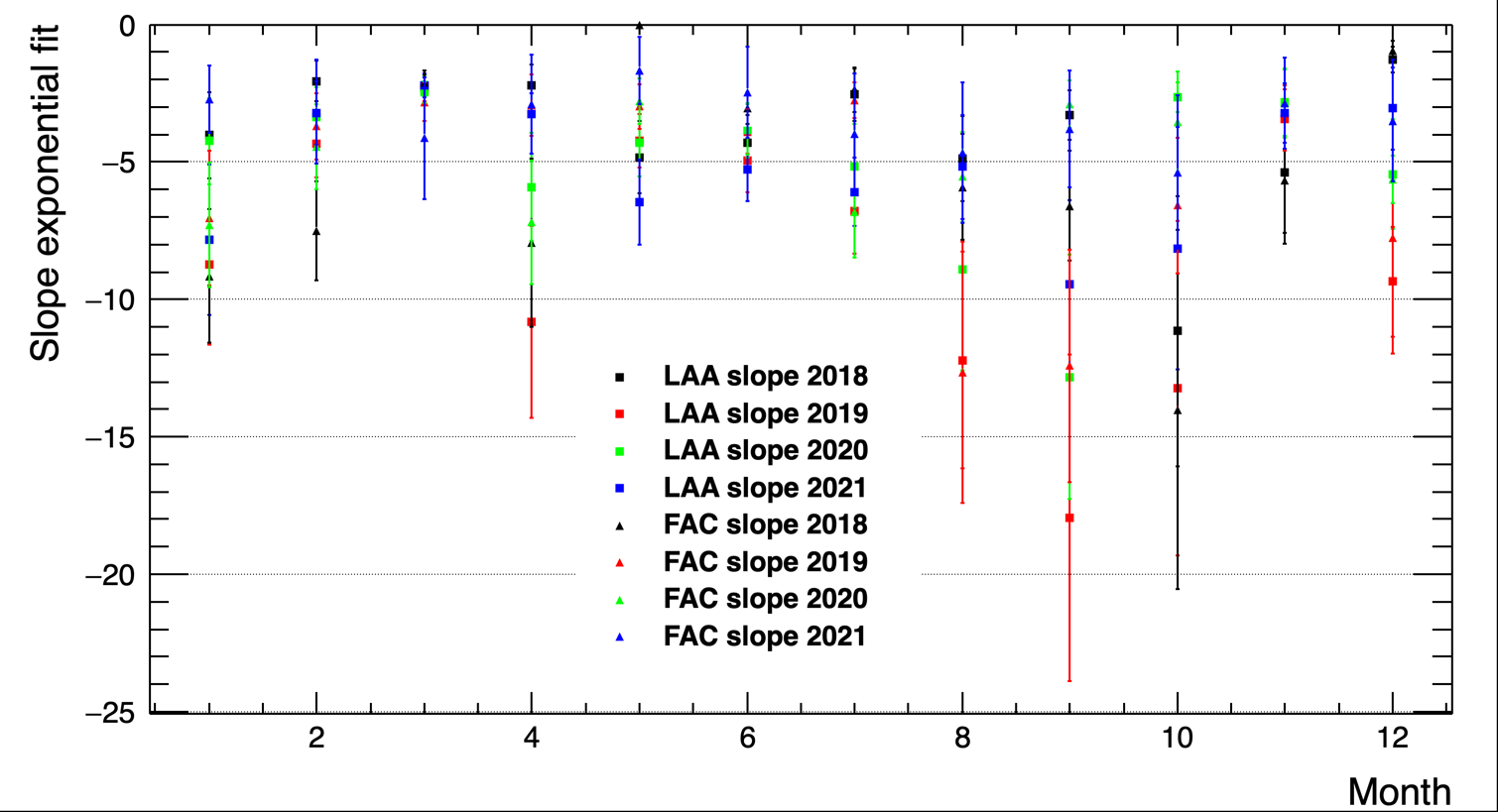}
    \caption{Slope exponential fit parameters for LAA and FAC stations.}
    \label{slope_fit}
\end{figure}

\begin{figure}
    \centering
    \includegraphics[width=0.6\linewidth]{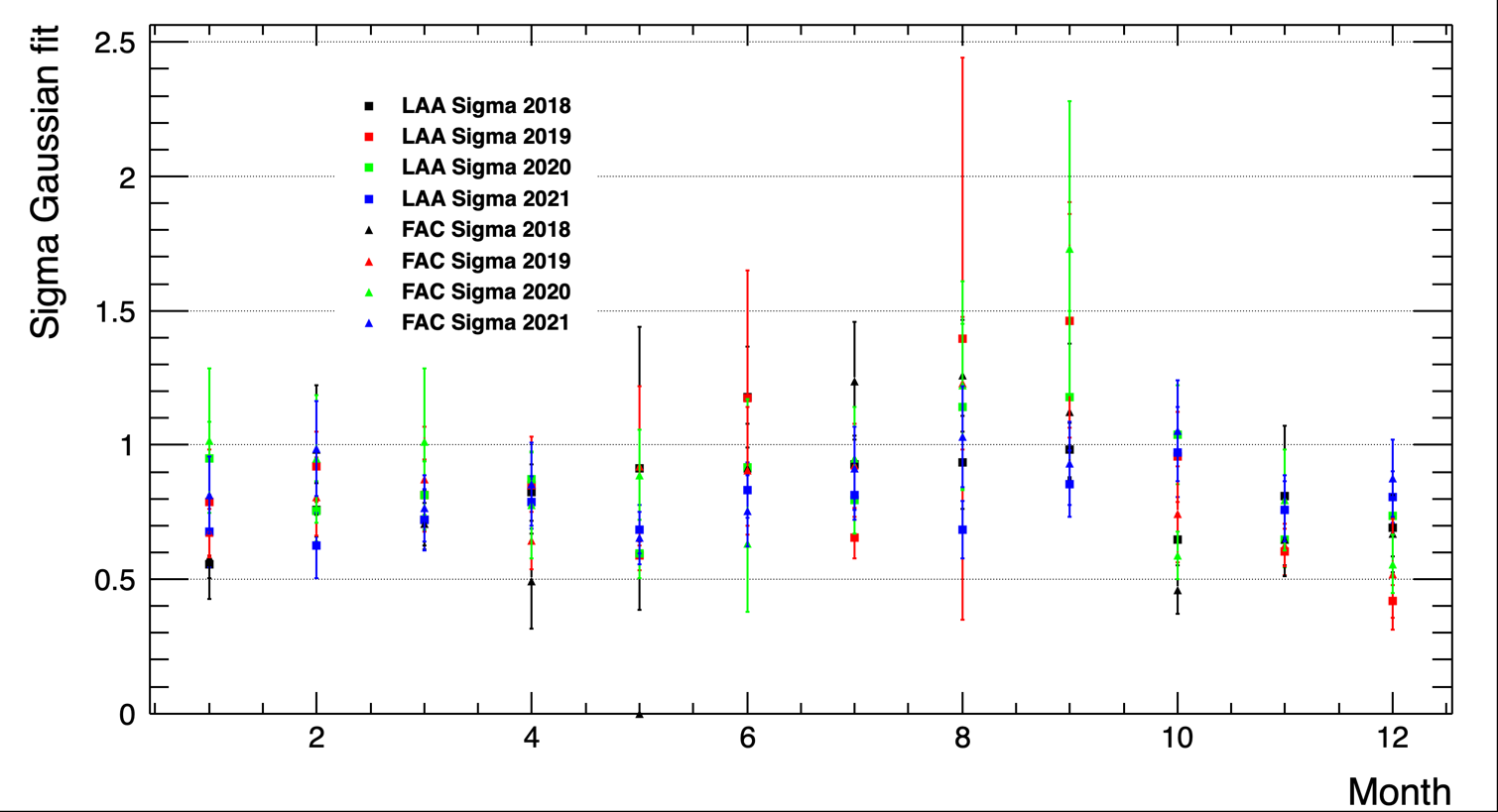}
     \includegraphics[width=0.6\linewidth]{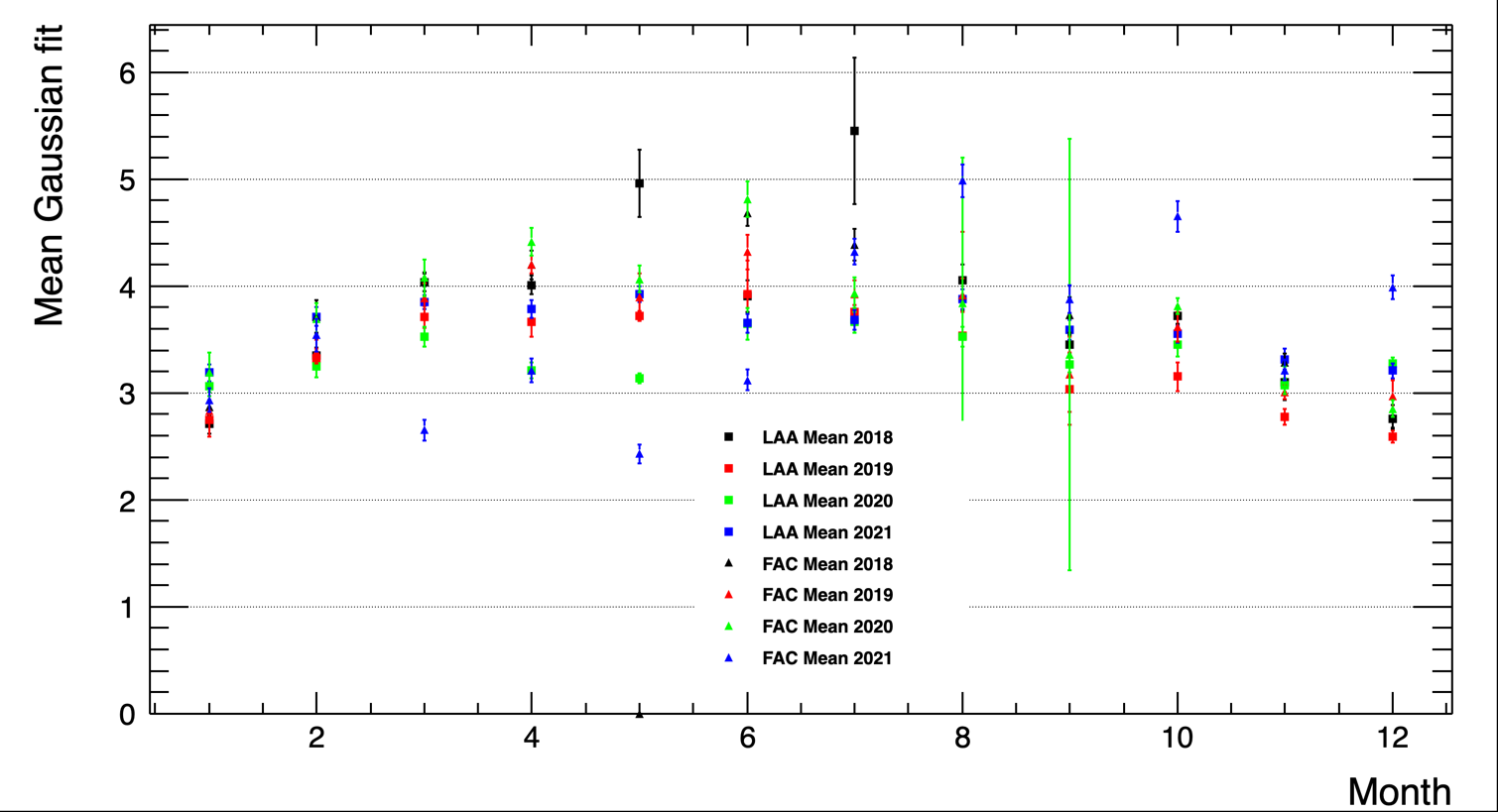}
    \caption{Sigma and mean Gaussian fit parameter distribution for LAA and FAC stations.}
    \label{sigma_mean_fit}
\end{figure}

\begin{figure}
    \centering
    \includegraphics[width=0.6\linewidth]{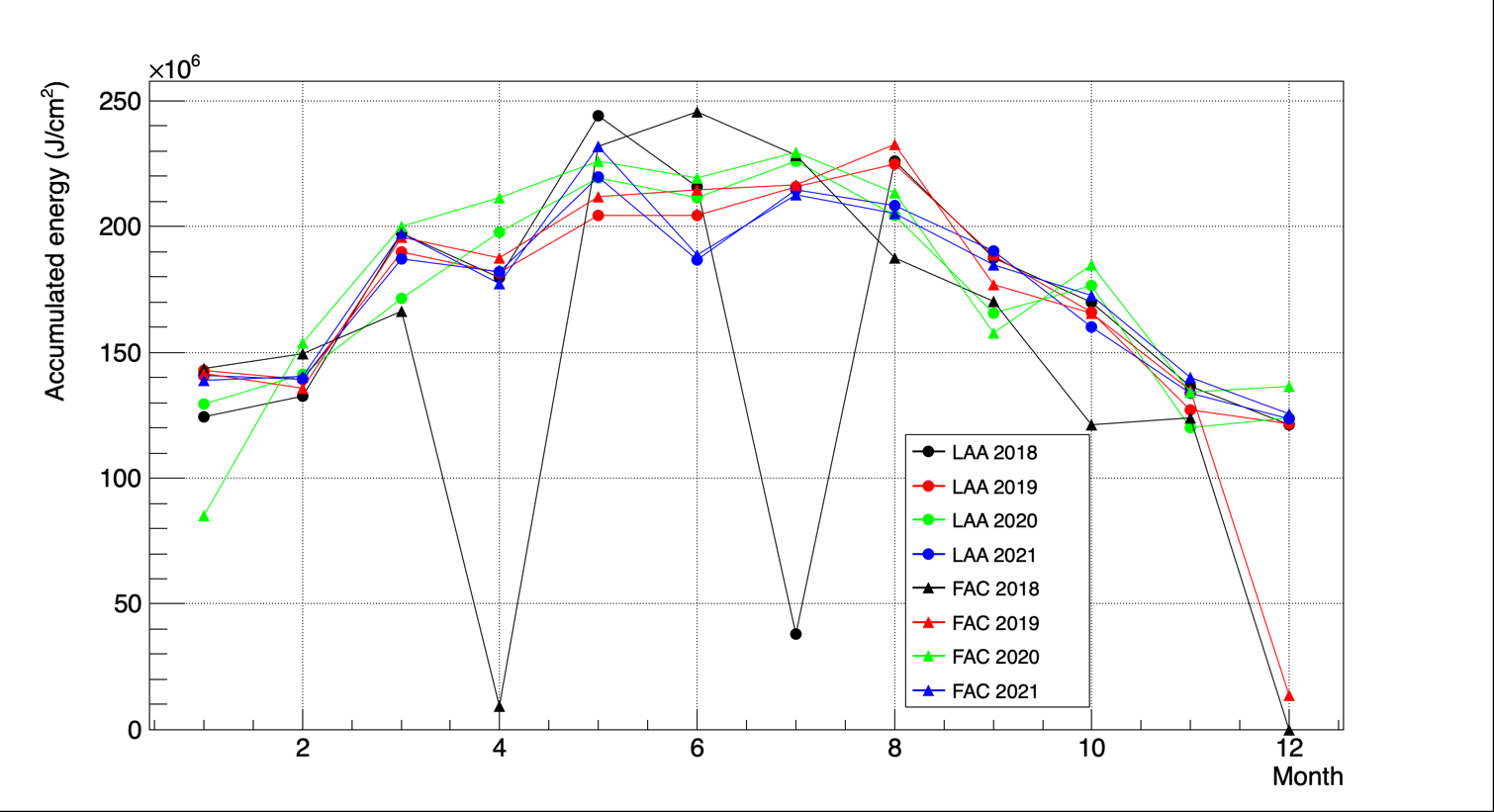}
    \caption{Total energy accumulated distribution for each month and the four years for LAA and FAAC stations.}
    \label{month_energy}
\end{figure}

It is possible to make a more detailed analysis considering the monthly SIV distribution. In Figure~\ref{His_jul_Jan}, as an example, it is shown the SIV distribution for the LAA station considering January and July in 2021. It can be appreciate that SIV are higher for July than for January, which corresponds to one of the months with the highest temperature (summer). These behavior is the same for other stations, resulting in an increase of SIV as the months approach summer and decreases as they approach to winter.\\ 
From these distributions, it is possible to obtain the mean and maximum average SIV, in order to show  variation in months. In Figure~\ref{mean_LAA_FAC}, as an example, it is shown these distributions for FAC and LAA stations, over all years. Clearly, mean average SIV are consistent with each other, also, maximum average SIV are consistent for each month. For this last distribution, it can be notice an increase and then a decrease as the SIV  approach  summer and move away as it approach winter. This increase is due (as it is well known) to the fact that in summer months the sun has a greater radiation on the Earth. The respective SIV measurements from other stations are consistent with these distributions. Therefore, the mean and maximum average SIV have not changed over the years.\\
To characterize the SIV distribution, it was made an exponential and Gaussian fit for the first and last part, respectively, as it is shown in Figure~\ref{His_jul_Jan} (green lines). Using the exponential fit slope parameter, the MPV drop can be characterized, while the maximum values of SIV distribution can be characterized by the sigma and the mean of the Gaussian fit, which represents the range and the mean of which they are found, respectively. With these three parameters a comparison between the SIV distributions can be made. For the intermediate distribution, it is not suitable for comparison due to the fluctuations that occur. For exemplary purpose, the fit parameter values are shown for LAA and FAC stations. The exponential slope fit parameters comparison are shown in Figure~\ref{slope_fit}, where it can be appreciate that they are consistent within error values, then, MPV slope has not changed. In Figure~\ref{sigma_mean_fit}  are shown the sigma and mean Gaussian fit parameters comparison, where they are clearly seen to be consistent. The large error bars for some parameters are due to the fact that in those months there were a few statistics to be able to make a good fit. A  Gaussian distribution can be appreciate for  maximum average value distribution in  Figure~\ref{mean_LAA_FAC}, with its maximum point located between the months of June and August, which are the summer months. Similar results are obtained for the other stations. Therefore, there is no change in SIV monthly-distributions. \\
To conclude this study, in Figure~\ref{month_energy} it is shown the total density energy (J/cm$^2$) accumulated in LAA and FAC stations. The values were obtained by integrating the SIV vs time graph. The low values are due to the null values,  for example, related to July 2018 only data was taken from 1 to 5 and from 24 to 31, i.e., only 13 days the LAA station took data on that day. All distributions have a Gaussian behavior, over again,  maximum value occurs during  summer. Same distributions are obtained for the other stations.

%Consistent results are obtained with the other stations, where it can be say that there has been no change in the SIV distributions.\\ %Then, to show a better appreciation for the results, the analysis for the LAA station is shown, because similar results are obtained for the other stations.\\

\subsection{Daily solar radiation study}\label{Daily}
\begin{figure}
    \centering
    \includegraphics[width=0.6\linewidth]{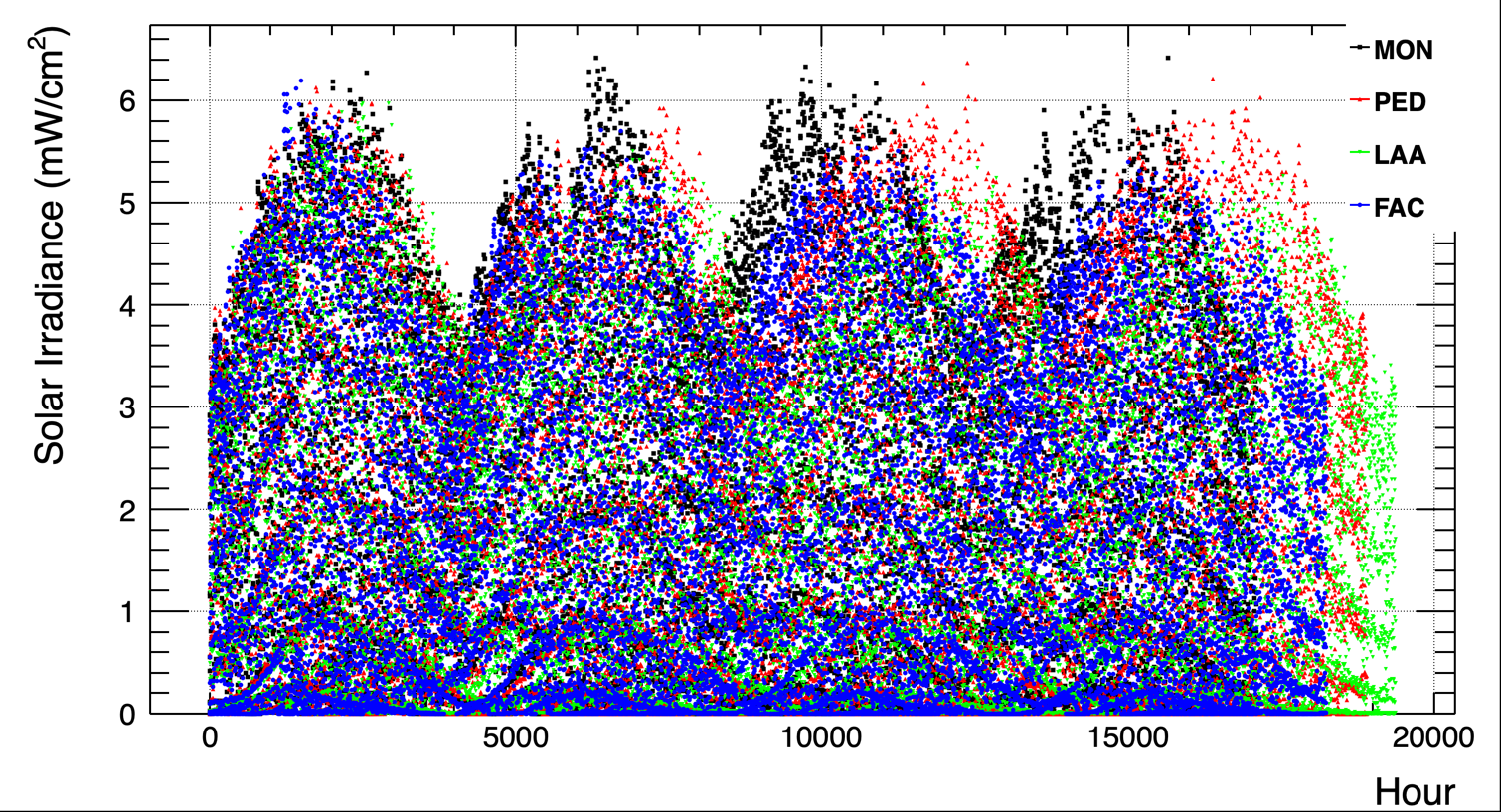}        \includegraphics[width=0.6\linewidth]{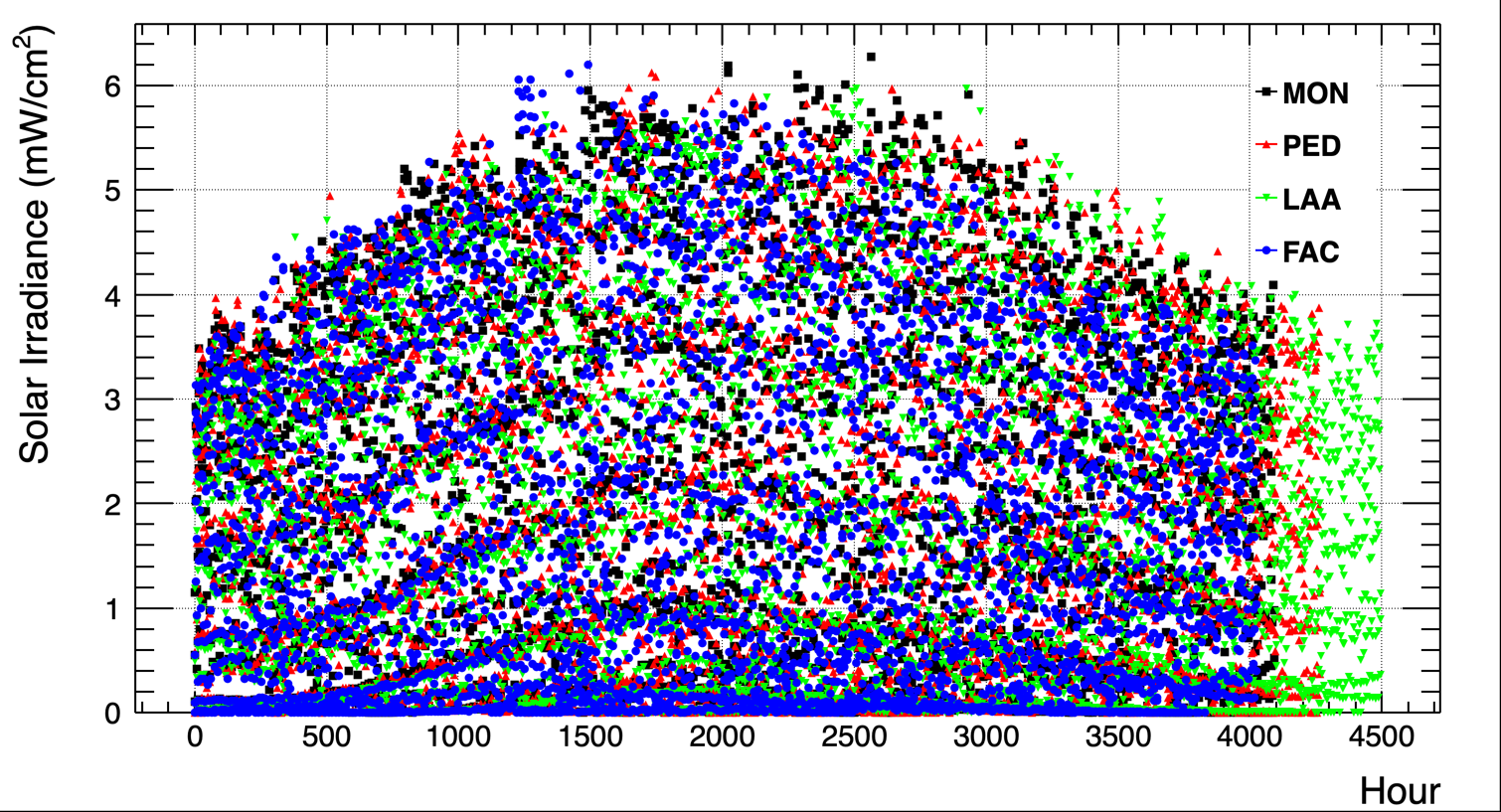}
    \caption{SIV as function of the number of hours during the four years. All non-null values are plotted (top panel). In the low panel it is shown the distribution for the year 2018.}
    \label{IrrVsHour}
\end{figure}

In Figure~\ref{IrrVsHour} the correlation between the SIVs with respect to the total number of hours in the four years is shown. In the top panel, the four lobes are associated with each year. For example, the first year distribution is shown in the low graph of Figure~\ref{IrrVsHour}, which is the distribution of 2018. 

\begin{figure}
    \centering
    \includegraphics[width=0.6\linewidth]{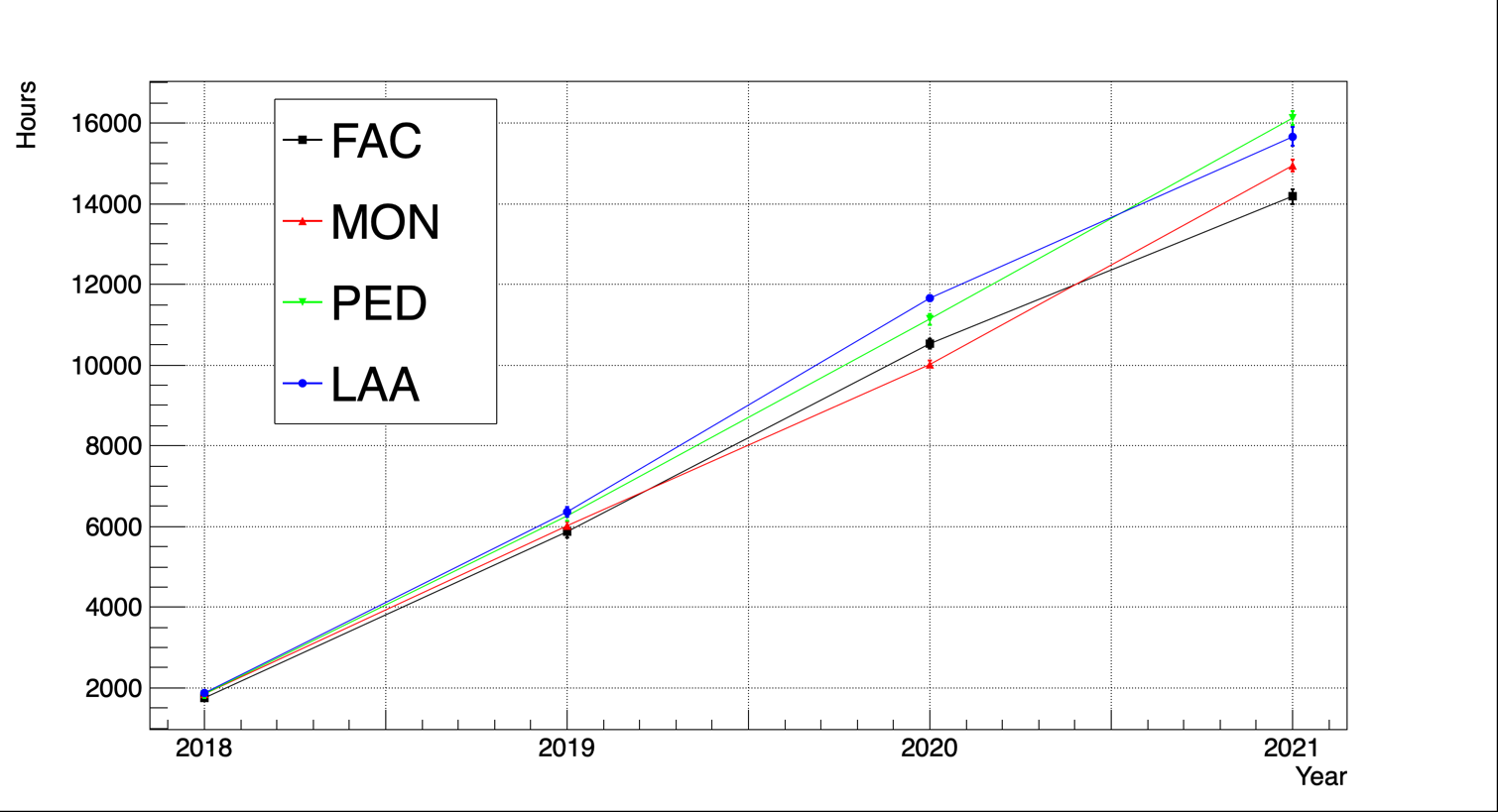}        \includegraphics[width=0.6\linewidth]{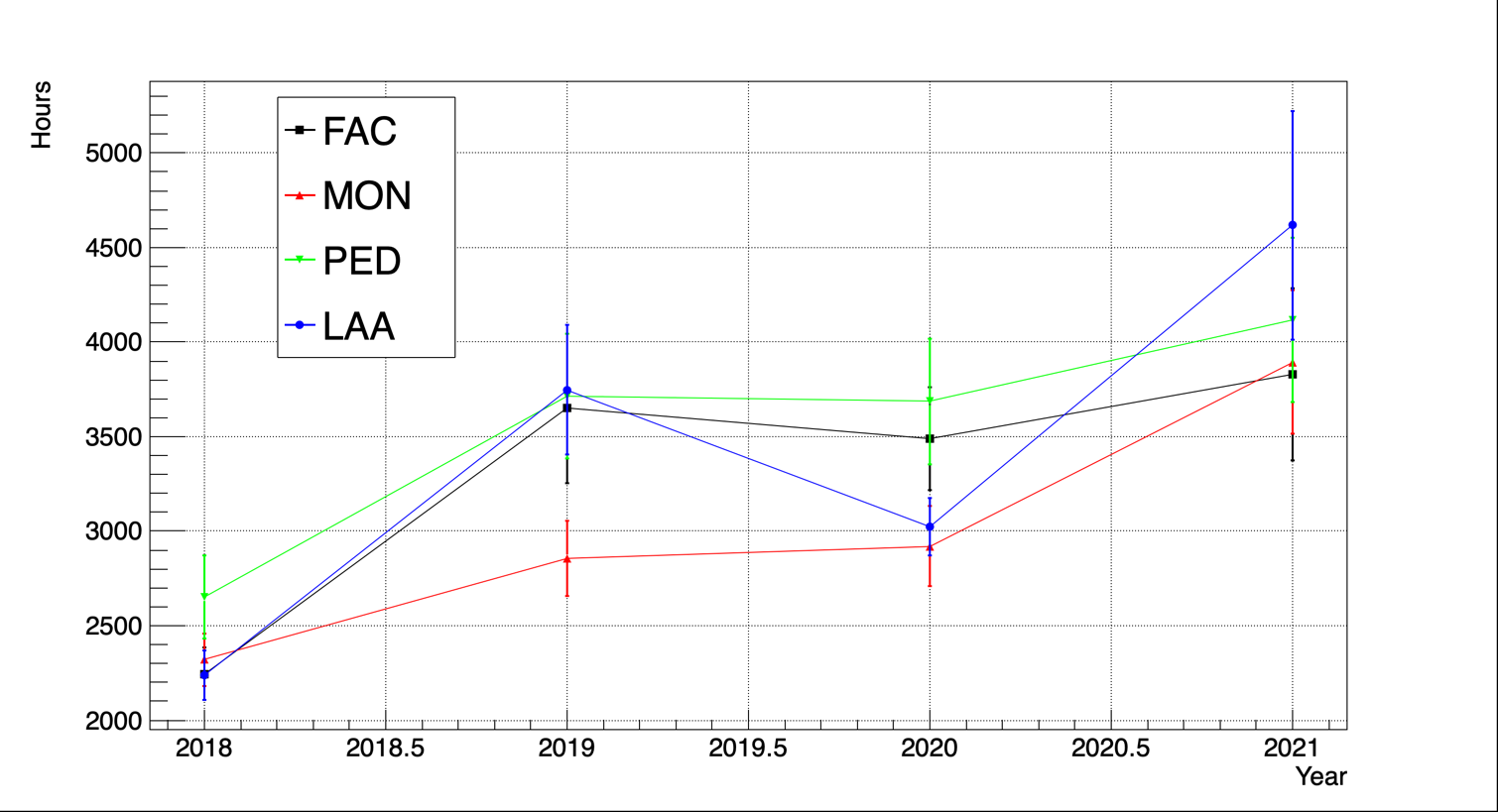}
    \caption{Mean (top) and sigma (down) fit parameter value from the Gaussian fit distribution obtained from Figure~\ref{IrrVsHour}.}
    \label{mean_sigma}
\end{figure}

%From the Figure~\ref{ratio_LAA_PED},  
This distributions could be compared by obtaining the Gaussian fit parameters (mean and sigma), which it was performed on each distribution by year. In Figure~\ref{mean_sigma}, it is shown the distribution of these parameters. The upper graph is the distribution of the means, it is rising in the hours due to the passage of the time. The bottom graph shows the sigma parameter distribution over the hours. %To have a better interpretation of the mean values, in Table~\ref{table_mean} it is shown the date interval in which the statistically mean value lies, this information was obtained by identifying the date with the respective time-hour value. 
The fluctuations in both distributions can be associated to the pollution, cloudiness and/or null-values. However, it is not a huge discrepancy and parameters can be considered consistent, within the error values. Then, there are no change in the SIV.
%\begin{table}[ht]
%\caption{Mean value interval day for each station per year (day/month).}
%\centering
%\begin{tabular}{ |c|c|c|c|c| } 
%\hline
%Station & 2018 & 2019 & 2020 & 2021 \\
%\hline
%MON & (01/06, 09/06) & (21/06, 30/06) & (26/06, 06/07)& (01/07, 15/07)\\ 
%PED & (15/06, 21/06) & (04/06, 03/07) & (31/06, 10/07)& (02/06, 15/06)\\ 
%LAA & (26/06, 02/07) & (16/06, 26/06) & (09/06, 15/06) & (22/06,10/07)\\
%FAC & (05/07, 11/07) & (17/06, 29/06) & (27/06, 07/07)& (28/06, 16/07)\\
%\hline
%\end{tabular}
%\label{table_mean}
%\end{table}

%As an example, the results will be displayed for the LAA station (as it was mentioned in the last section) and at the end of this section, the results of the other stations will be displayed, in order to compare them.
\begin{figure}
    \centering
    \includegraphics[width=0.6\linewidth]{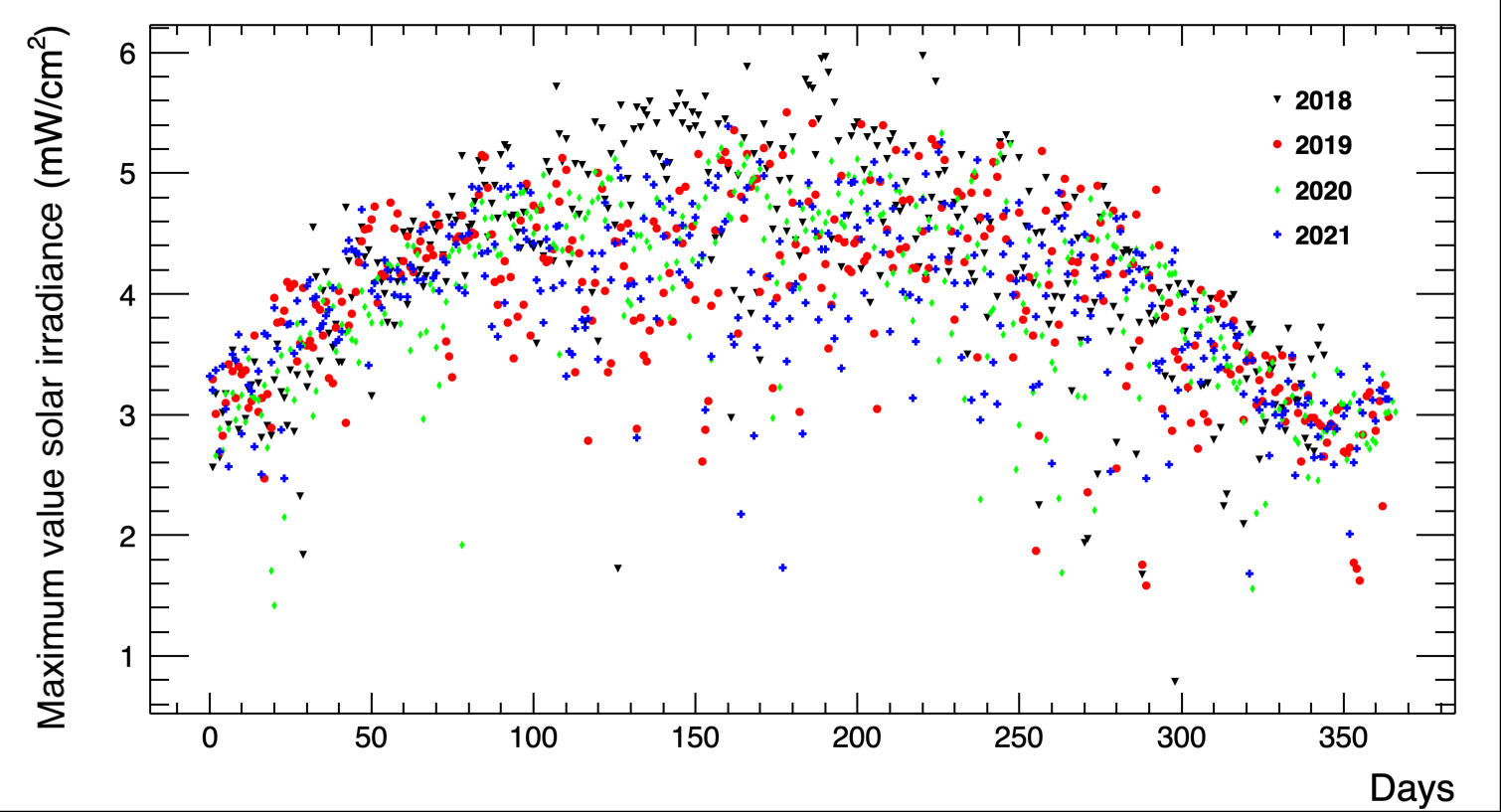}        \includegraphics[width=0.6\linewidth]{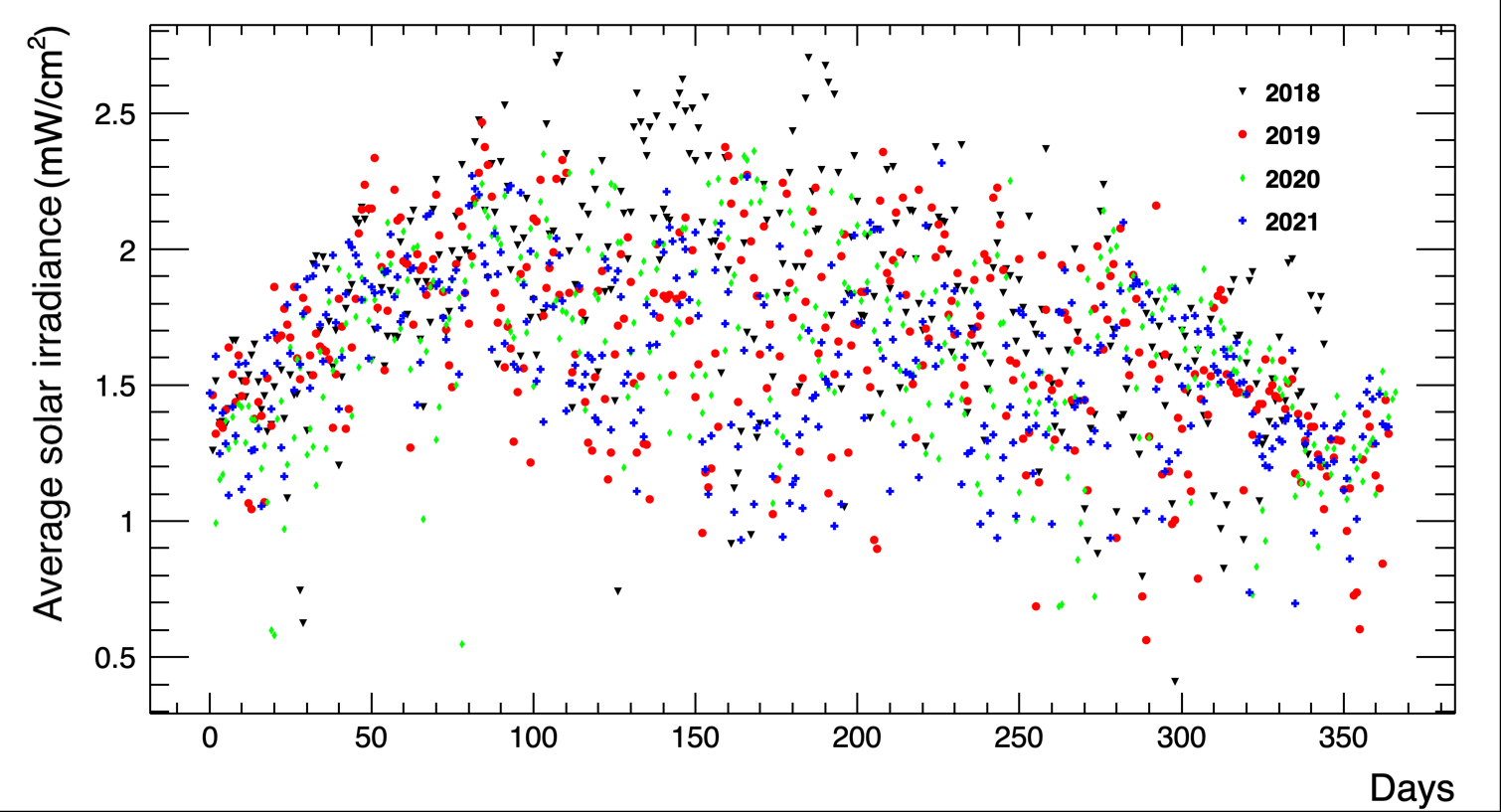}
    \caption{Maximum SIV (top) and average SIV (down) per day for all years for the LAA station.}
    \label{ave_max}
\end{figure}

\begin{figure}
    \centering
    \includegraphics[width=0.6\linewidth]{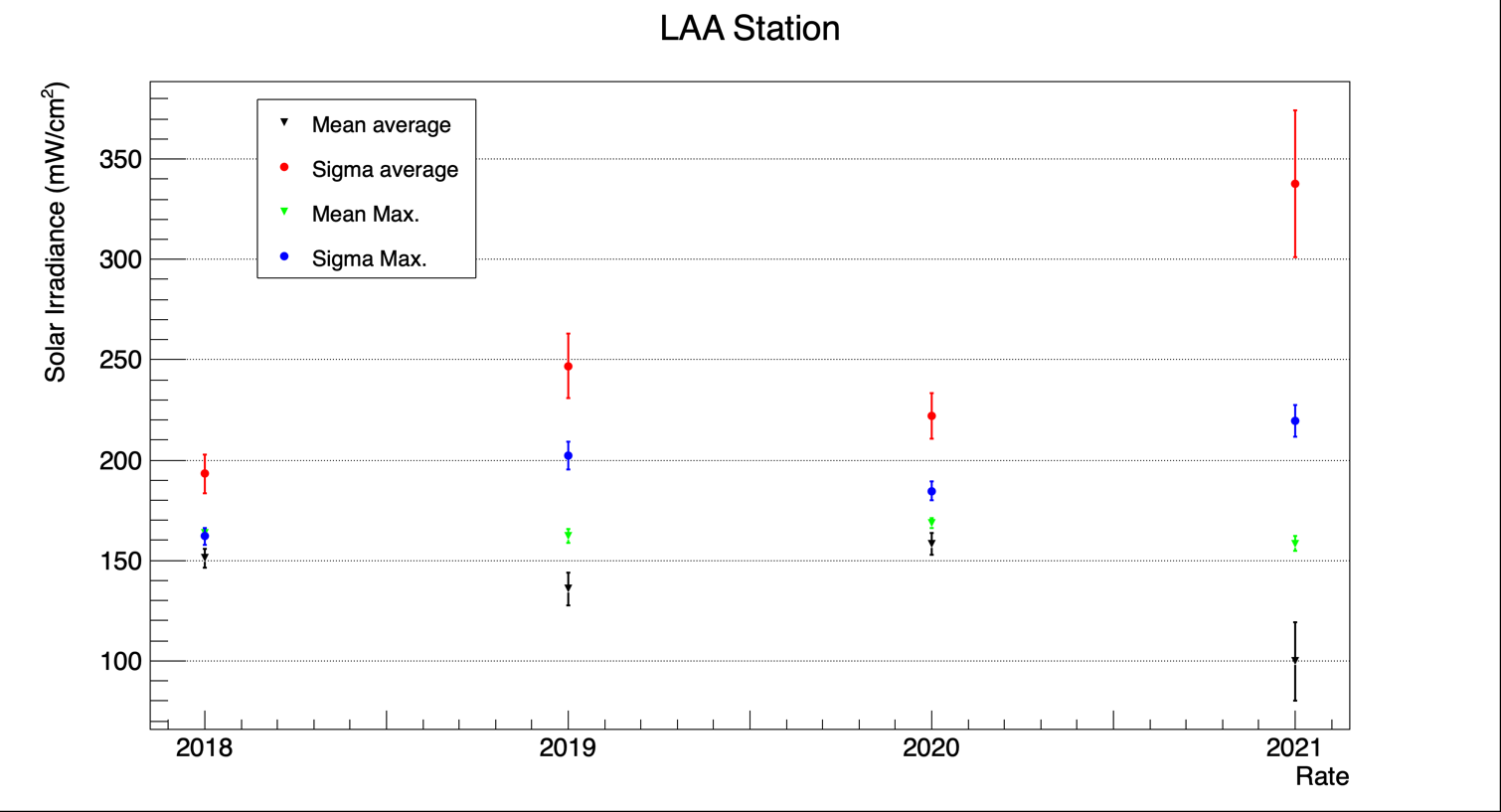}
    \caption{Mean and sigma parameters from the Gaussian fit of Figure~\ref{ave_max}.}
    \label{mean_sigma_average}
\end{figure}

\begin{figure}
    \centering
    \includegraphics[width=0.6\linewidth]{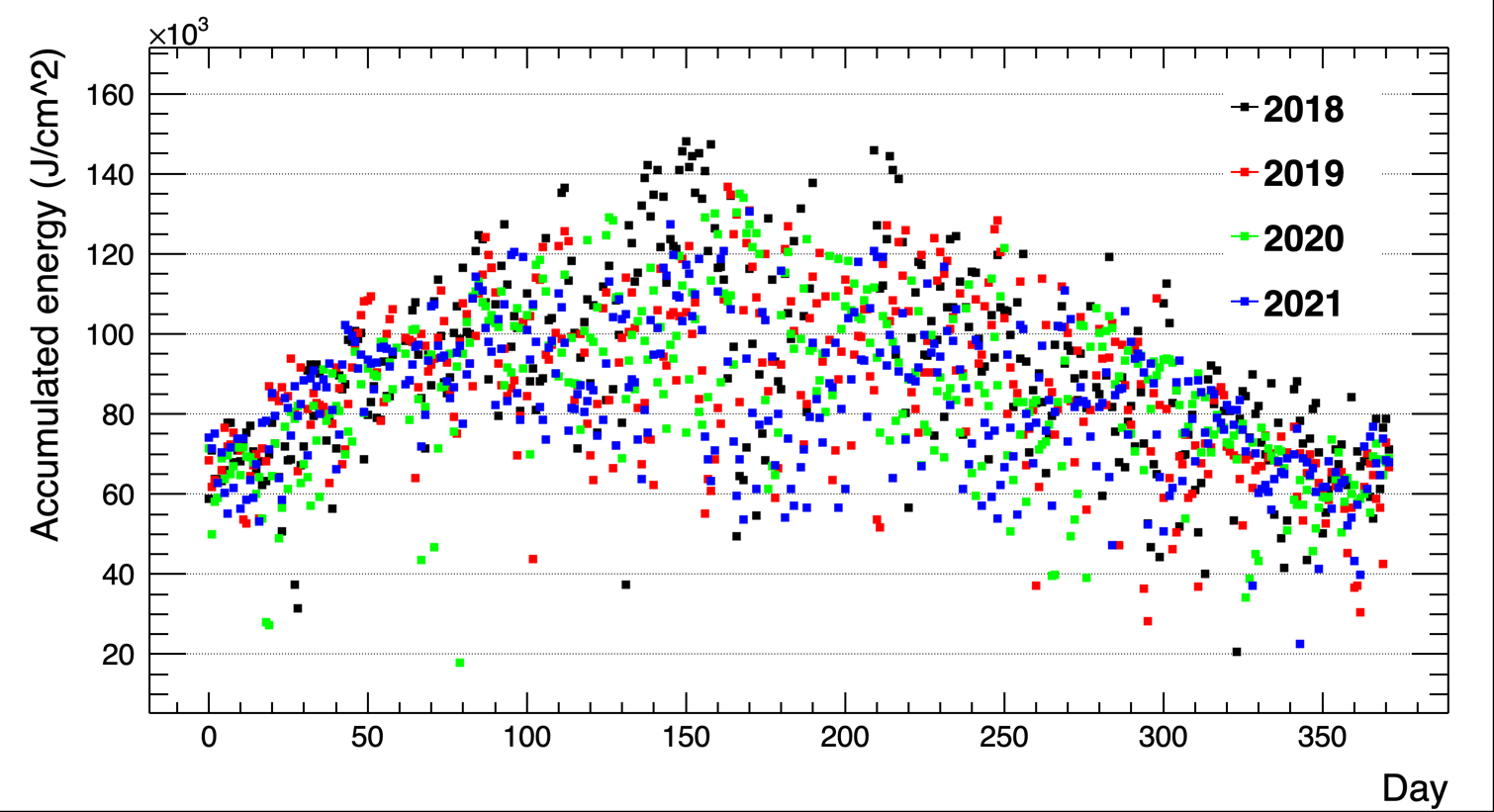}
    \caption{Energy density distribution accumulated in LAA station from 2018 to 2021.}
    \label{density_day}
\end{figure}

From Figure~\ref{IrrVsHour}, it was obtained the SIV average and the maximum SIV average for every day and for LAA station (as an example), these distributions are shown in Figure~\ref{ave_max}. In order to compare these distributions, a Gaussian fit was performed to obtain the mean and sigma values, then, each year can be characterized by these parameters. These parameters are compared in Figure~\ref{mean_sigma_average}. The maximum average value (as expected) is greater than average value, while  sigma parameter of  SIV average is greater than the respective maximum SIV average, this behavior is due to fluctuations, as it was mentioned before.\\
The energy density distribution accumulated could be obtained for all stations. As an example, in Figure~\ref{density_day} it is shown the density distribution for LAA station from 2018 to 2021. Similar distributions were obtained for the other stations, having a grater value during the summer months, as expected. The observed fluctuations are due to the null values.

%that the average value of mean has greater fluctuations than the maximum SIV, as it was mentioned before.%, as it can be appreciate in Figure~\ref{ave_max}.

\begin{figure}
    \centering
    \includegraphics[width=0.6\linewidth]{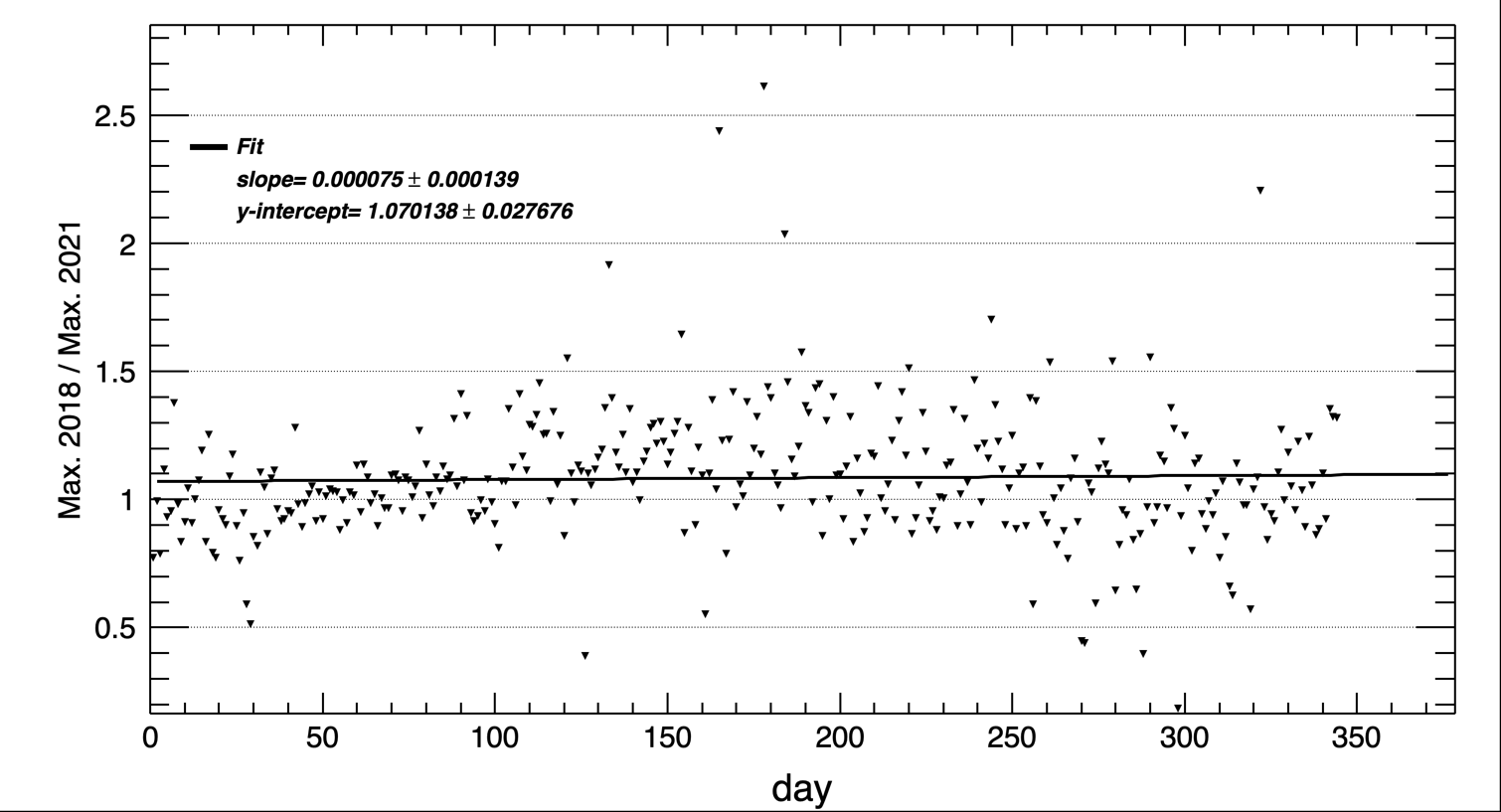}
        \includegraphics[width=0.6\linewidth]{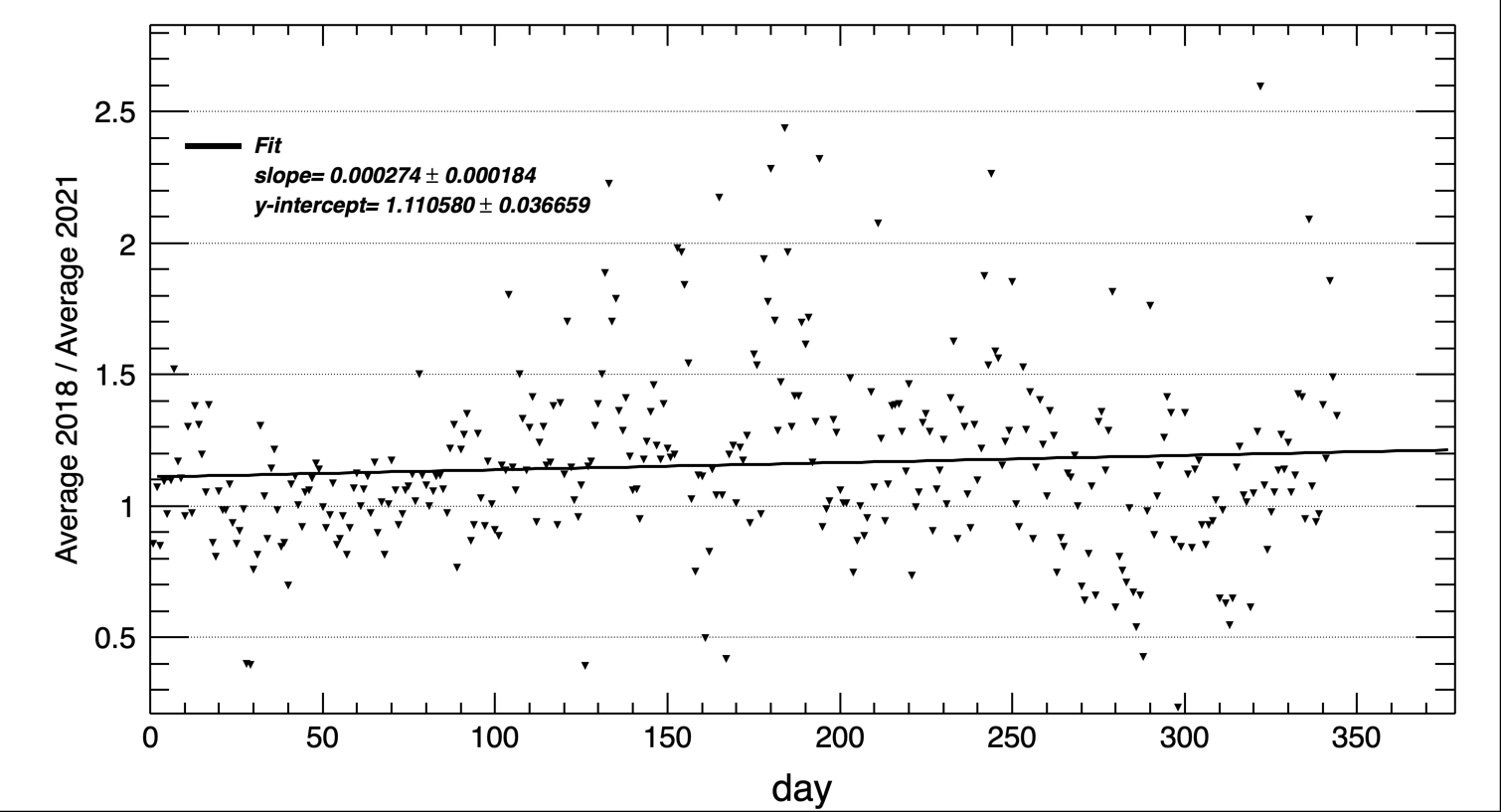}
    \caption{Distribution of the SIV ratio for the maximum (top) and average (down) per day.}
    \label{slope}
\end{figure}

\begin{figure}
    \centering
    \includegraphics[width=0.6\linewidth]{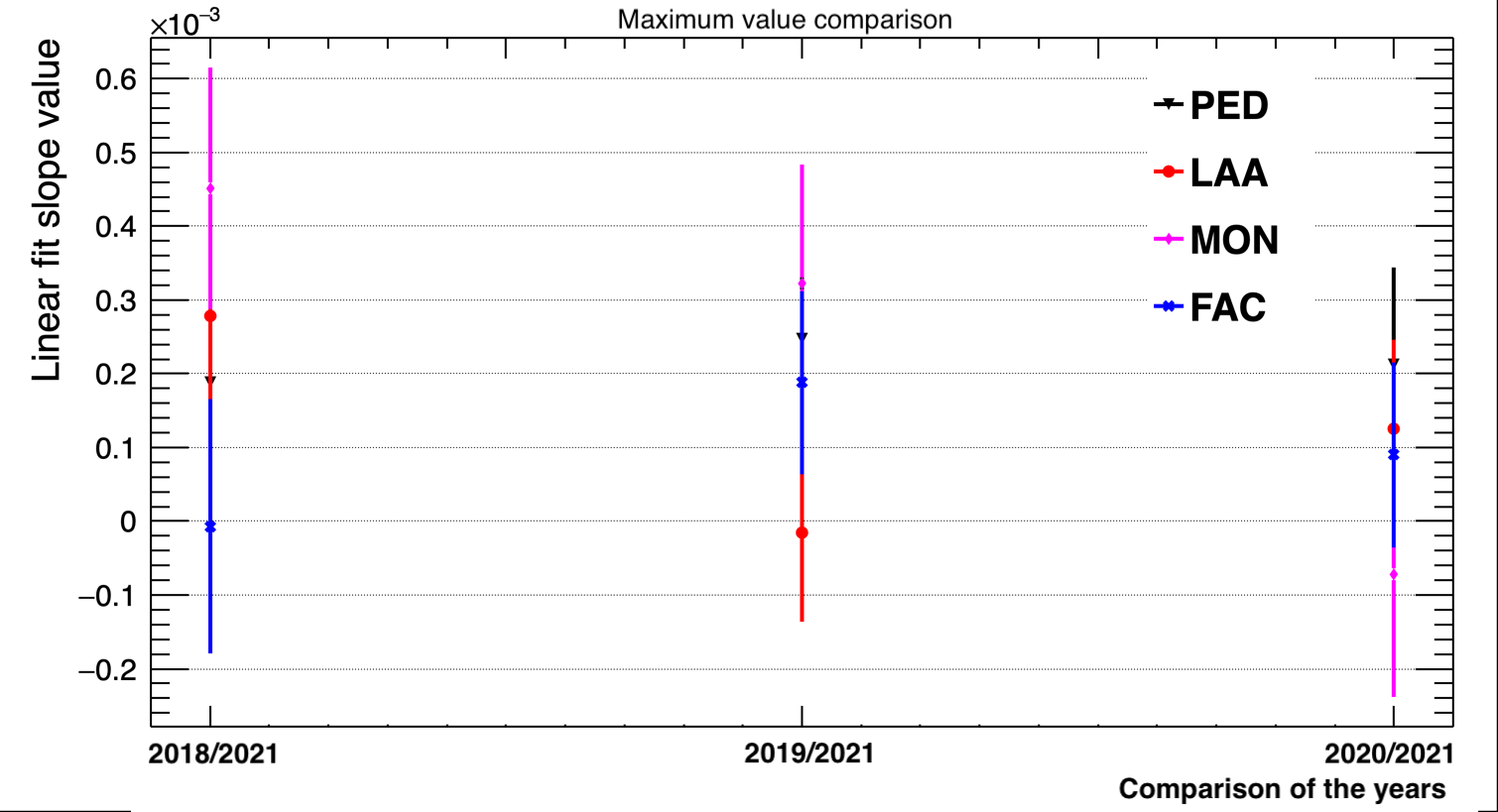}
    \includegraphics[width=0.6\linewidth]{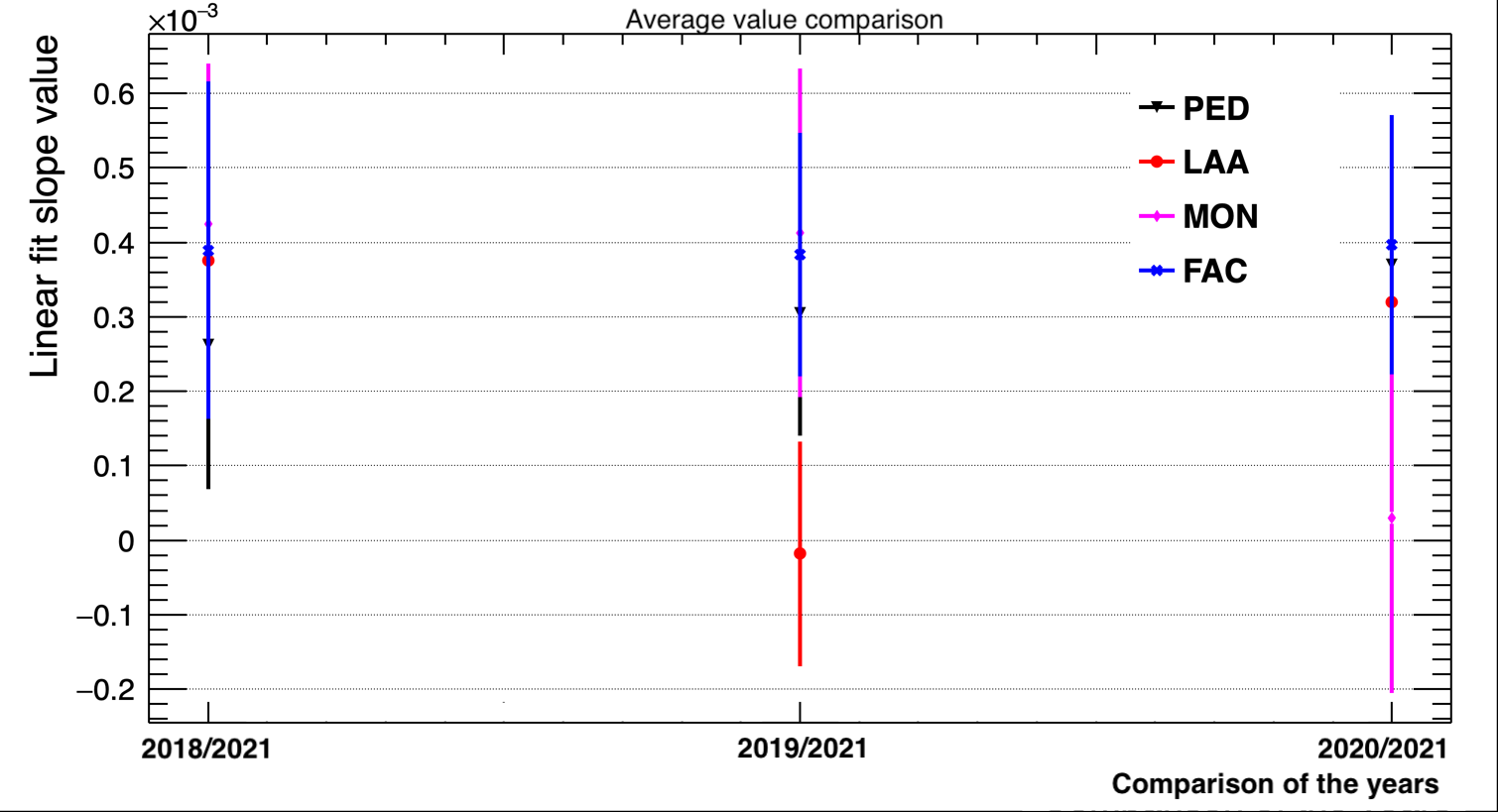}    
    \caption{Comparison of slope parameter for maximum of SIV  (top) and for average SIV (dow) per day.}
    \label{slope_average}
\end{figure}

A finally analysis could be done by comparison of values per day from Figure~\ref{ave_max}, where the ratio of the daily SIV 2018, 2019 and 2020 with respect to 2021 was considered, for each distribution. In Figure~\ref{slope} it is shown the rate for the maximum and average values per day, i.e, for each day it was calculated the rate maximum and the average SIV, a linear fit is also shown. Slope parameter value of linear fit is the most important parameter of these fit distribution, since this shows an increase or decrease in the SIV comparison. It is seen that for the case of maximum SIV is consistent with zero, which means that there has not been any variation in these years. %For the case of the average SIV, due to the fluctuation of this value, the slope can be not consistent with zero, however, it is very small.
Comparison of slope parameter for maximum  and average SIV are shown in Figure~\ref{slope_average}, as it can be seen, the values are consistent with zero, therefore SIV has not changed from 2018 to 2021. %the angle associated with the linear fit is at least of the order of $\sim$0.15$^o$. Same analysis was made for the average values, in which it is obtained that the slope is consistent with the value of one.\\
%Similar results has been obtained for the other years and stations, where it is obtained the same tendency.
 
\begin{center}
Acknowledgments   
\end{center}
 The authors would like to thank the \textit{Secretar\'ia del Medio Ambiente de la Ciudad de M\'exico (SEDEMA)}, for providing the data in a public manner, in order to perform this work. As well as thanking Mr. Melit\'on Bustamante Pel\'aez in explaining the operation of the actinograph and data tacking.   
\section{Discussion and conclusions}~\label{Discussions}
In this work, a statistical analysis was carried out for the study of solar radiation in CDMX from 2018 to 2021. The data was taken by several weather stations, which, every hour taken measurements during the years period, according to~\cite{base}. Then, analysis could be performed by years, months and days.\\
The annual analysis turned out to have various values fluctuations in the graphs, due to null SIV. However, it was possible to perform an analysis with the average and maximum SIV per years, where it was shown a SIV for the maximum value in a range (5.40, 6.42) mW/cm$^2$ and for average SIV (1.7, 2.3) mW/cm$^2$. The MPV SIVs are consistent for all stations and years, which only has a variation of 0.0010~mW/cm$^2$, as shown in  Figures~\ref{mpvvalueperyear} for all stations. With these distributions, it can be concluded there is no change of SIV over the years.\\ 
From the monthly analysis it was possible to characterize the SIV distributions from 2018 to 2021, when performing an exponential fit to the MPV and a Gaussian fit to the final part of the distribution, which represents the maximum values distribution (Figure~\ref{His_jul_Jan}). From these parameters, it is conclude, all stations measured the same SIV, because there has been no change within the statistical errors (Figures~\ref{slope_fit} and \ref{sigma_mean_fit}). A final analysis could be performed by calculating the total density energy (mW/cm$^2$) received in each station, which, as it can be expected, the maximum energy occurs during the summer. Then, the SIV has remained unchanged.\\
Finally, a more detailed analysis was performed to consider the days SIV, in such a way that statistical parameters can be found to characterize the distributions considering the fluctuations. It was obtained the correlation between SIV and respective hour (Figure~\ref{IrrVsHour}), where a Gaussian fit was made to obtain the sigma and mean Gaussian fit parameters , which are consistent, as it was shown in Figure~\ref{mean_sigma}. In addition, it was obtained the maximum and average SIV per, an example was shown in Figure~\ref{mean_sigma} for LAA station, where fit parameters values are consistent with each other. In addition it was obtained the mean average and maximum SIV from distributions of Figure~\ref{IrrVsHour} per day, as an example in Figure~\ref{ave_max} it is shown  for LAA station. By performing a Gaussian fit, it could be obtained a comparison of fit parameters as it was shown in Figure~\ref{mean_sigma_average}. Where it was obtained that the mean of the maximum SIV distribution has been invariant, within its error values, which means that the mean of the maximum values has not changed. However, the mean average SIV has had fluctuations due to the fluctuations of the SIV, which are associated with population, cloudiness, etc. The sigma parameter for maximum values is less than the average value, due to the mentioned fluctuations. The mean values of these distributions have been consistent. Finally, a comparison was made by dividing the SIV per day from 20118, 2019, 2020 with respect to the 2021. When performing a linear fit, it was observed that the slope parameter is consistent with zero, as it is shown in Figure~\ref{slope_average}. Therefore, the SIV has been not changed.\\
From these analyses it has been shown that the SIV have remained constant over the period 2018-2021 in CDMX, which is agree with a previously studies in various part of the world~\cite{model,model2,article_irradiance1,article_irradiance2}. However, this constant value does not imply that it does not cause damage to organism as already mentioned~\cite{dermatologia, secretariasalud, gallegos, economista, mata}, because health damage is due to exposure time.

\end{document}